\documentclass[nofootinbib,english,notitlepage,superscriptaddress,showkeys,preprintnumbers,longbibliography]{revtex4-1}

\pdfoutput=1

\usepackage[T1]{fontenc}
\usepackage[latin9]{inputenc}
\usepackage{babel}
\usepackage{amsmath}
\usepackage{amssymb}
\usepackage{wasysym}
\usepackage{graphicx}
\usepackage{esint}
\usepackage[unicode=true,
 bookmarks=false,
 breaklinks=false,pdfborder={0 0 1},backref=false,colorlinks=true,citecolor=blue,urlcolor=blue,linkcolor=blue]
 {hyperref}

\makeatletter

\@ifundefined{textcolor}{}
{%
 \definecolor{BLACK}{gray}{0}
 \definecolor{WHITE}{gray}{1}
 \definecolor{RED}{rgb}{1,0,0}
 \definecolor{GREEN}{rgb}{0,1,0}
 \definecolor{BLUE}{rgb}{0,0,1}
 \definecolor{CYAN}{cmyk}{1,0,0,0}
 \definecolor{MAGENTA}{cmyk}{0,1,0,0}
 \definecolor{YELLOW}{cmyk}{0,0,1,0}
}

\usepackage{hyperref}
\usepackage{bbm}
\usepackage{cleveref}
\usepackage{bbold}

\usepackage{feynmp}
\usepackage{todonotes}
\definecolor{lightblue}{HTML}{A9D0F5}
\definecolor{lightgreen}{HTML}{BCF5A9}
\definecolor{lightred}{HTML}{F6CECE}
\definecolor{lightorange}{HTML}{FFA800}

\definecolor{greengray}{HTML}{5C9393}
\definecolor{lightgreengray}{HTML}{80CCCC}

\newcommand{\td}{\text{d}}
\newcommand{\Mp}{M_\text{P}}
\newcommand{\vac}{\text{vac}}
\newcommand{\Lambdamax}{\Lambda^{(\text{max})}_{\text{LB}}}

\newcommand*\imgghostsmall{\vcenter{\hbox{\includegraphics[width=6px, height=7px]{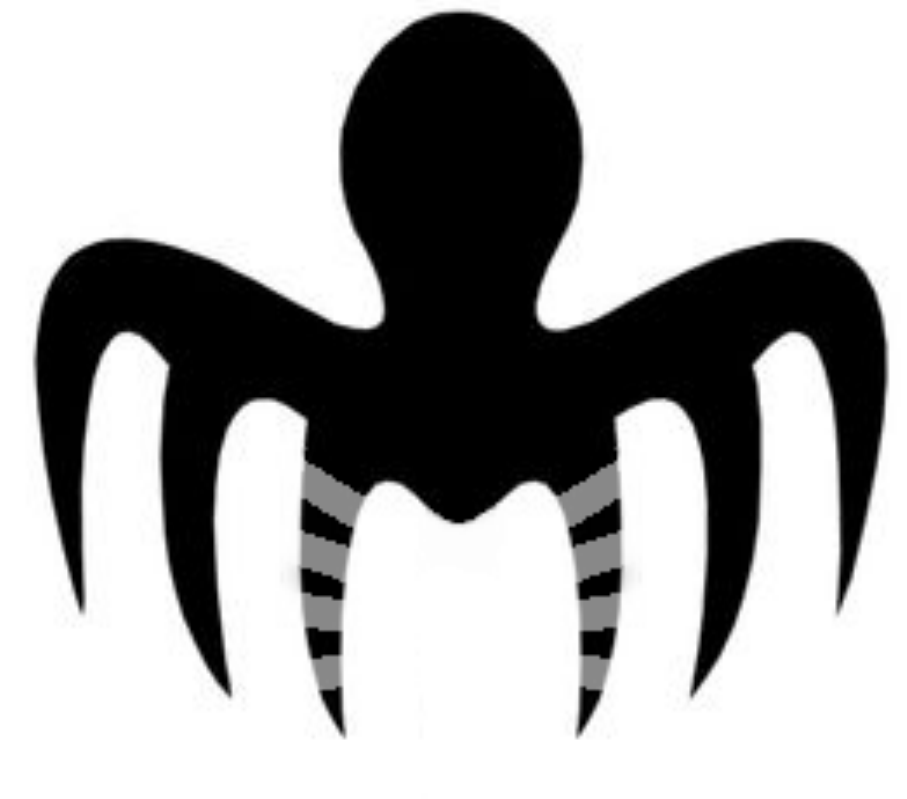}}}}
\newcommand*\imgghosttiny{\vcenter{\hbox{\includegraphics[width=4px, height=5px]{img/ghost_symbol.pdf}}}}

\newcommand*\ghost{\Phi_{\imgghostsmall}}
\newcommand*\ghostsmall{\Phi_{\imgghosttiny}}

\begin{document}

\preprint{NORDITA-2016-53}

\title{A spectre is haunting the cosmos:\\
Quantum stability of massive gravity with ghosts}

\author{Frank K\"{o}nnig}

\email{koennig@thphys.uni-heidelberg.de}

\selectlanguage{english}%

\affiliation{Institut f\"{u}r Theoretische Physik, Ruprecht-Karls-Universit\"{a}t Heidelberg,
Philosophenweg 16, 69120 Heidelberg, Germany}

\author{Henrik Nersisyan}

\email{nersisyan@thphys.uni-heidelberg.de}

\selectlanguage{english}%

\affiliation{Institut f\"{u}r Theoretische Physik, Ruprecht-Karls-Universit\"{a}t Heidelberg,
Philosophenweg 16, 69120 Heidelberg, Germany}

\author{Yashar Akrami}

\email{akrami@thphys.uni-heidelberg.de}

\selectlanguage{english}%

\affiliation{Institut f\"{u}r Theoretische Physik, Ruprecht-Karls-Universit\"{a}t Heidelberg,
Philosophenweg 16, 69120 Heidelberg, Germany}

\author{Luca Amendola}

\email{amendola@thphys.uni-heidelberg.de}

\selectlanguage{english}%

\affiliation{Institut f\"{u}r Theoretische Physik, Ruprecht-Karls-Universit\"{a}t Heidelberg,
Philosophenweg 16, 69120 Heidelberg, Germany}

\author{Miguel Zumalac\'{a}rregui}

\email{miguel.zumalacarregui@nordita.org}

\selectlanguage{english}%

\affiliation{Nordita, KTH Royal Institute of Technology and Stockholm University,
Roslagstullsbacken 23, 10691 Stockholm, Sweden}

\affiliation{Institut f\"{u}r Theoretische Physik, Ruprecht-Karls-Universit\"{a}t Heidelberg,
Philosophenweg 16, 69120 Heidelberg, Germany}
\begin{abstract}
Many theories of modified gravity with higher order derivatives are
usually ignored because of serious problems that appear due to
an additional ghost degree of freedom. Most dangerously, it causes an
immediate decay of the vacuum. However, breaking Lorentz invariance
can cure such abominable behavior. By analyzing a model that describes
a massive graviton together with a remaining Boulware-Deser ghost
mode we show that even ghostly theories of modified gravity can yield
models that are viable at both classical and quantum levels and, therefore,
they should not generally be ruled out. Furthermore, we identify
the most dangerous quantum scattering process that has the main impact
on the decay time and find differences to simple theories that only
describe an ordinary scalar field and a ghost. Additionally, constraints
on the parameters of the theory including some upper bounds on the Lorentz-breaking
cutoff scale are presented. In particular, for a simple theory of massive gravity we find that a breaking of Lorentz invariance is allowed to happen even at scales above the Planck mass. Finally, we discuss the relevance
to other theories of modified gravity.
\end{abstract}

\keywords{modified gravity, massive gravity, ghost, quantum stability, vacuum decay, dark energy}

\date{\today}

\maketitle

\begin{quote}{\it This work is dedicated to our friend, Tham.}
\end{quote}

\section{Introduction}

Despite the fact that the standard theory of gravity, general relativity (GR),
is already around one century old, there has always been interest
in finding viable modifications to it. In particular, the discovery of the late-time acceleration
of our Universe \cite{Riess:1998cb,Perlmutter:1998np} driven by some dark energy has
led to an additional motivation, as GR requires a technically unnatural
cosmological constant (CC) in order to be compatible with observations.
The list of problems with the standard theory goes much further (see for example Ref.~\cite{Bull:2015stt} for a recent review): GR
is not renormalizable and can only be regarded as an effective field
theory (EFT). Furthermore, the formation of structure at early times needs
an additional inflationary epoch, and even the requirement of some additional
dark matter might be the consequence of the inability of GR to properly describe
the evolution of the cosmic structure. It is however not only these
problems that make a search for modifications of GR attractive; there
is also the more fundamental question of which classes of theories are
allowed and consistent.

Under certain assumptions, Vermeil and Cartan independently proved
that Einstein equations are the only allowed field equations to describe gravity \cite{Vermeil1917,Cartan1922}.
In particular, if a rank-2 tensor $K$ is naturally constructed from
only a pseudo-Riemannian metric $g$, and is symmetric, divergence-free,
only second order in the derivatives of $g$, and linear in these derivatives,
then $K$ has to be a linear combination of the Einstein tensor and
the metric itself. Later, Lovelock showed that the requirements of symmetry
and linearity are redundant in four dimensions \cite{Lovelock1972}.
A generalization of this theorem has recently been suggested by Navarro
and Sancho \cite{NavarroSancho2007}, who replaced the assumptions
of the number of dimensions and absence of higher order derivatives
by a simpler requirement that $K$ is homogeneous, i.e., $K(\lambda^2 g) = \lambda^w K(g)$ $\forall g$, $\forall \lambda > 0$, and of weight $w > -2$.

Thinking about modifying GR can be translated into relaxing these
so-called Lovelock assumptions. Higher dimensional spacetimes, as well
as unnaturalness (understood in the mathematical sense, i.e., either breaking of locality or generally
non-$\mathcal{C}^{\infty}$), enable a richer phenomenology and do
not necessarily require a CC in order to fit current observations.
Additionally, a pseudo-Riemannian geometry is quite restrictive as
it both implies a vanishing torsion and enforces the connection to
be metric-compatible. Certainly the strongest assumption, however, is
the dependence on the metric only. Consequently, most theories of
modified gravity assume additional fields that can be either scalar,
vector, or tensor.

There is, however, one assumption that usually stays untouched: the
absence of higher order derivatives. An old theorem from Ostrogradski
states that non-degenerate Lagrangians that lead to third or higher
order derivatives in the equations of motion (EoM) always house an
additional ghost, i.e., a degree of freedom with the wrong kinetic sign.
But even degenerate Lagrangians producing third order derivatives
are affected by ghosts \cite{MotohashiSuyama2014}. The consequences
that come along with a ghost are usually believed to be fatal (see, e.g., Refs. \cite{Twain1875,Woodard2006}). Such a negative energy
mode could drive the classical theory into an instability. Even though
this might still be acceptable as long as the theory is in agreement
with observations, it indeed limits the number of viable theories
drastically. The real catastrophe appears, however, at the quantum level:
ghost fields can decay into ordinary matter fields by reaching
arbitrarily large negative energy states. And, even worse, this decay
will practically happen instantaneously (see Refs. \cite{Woodard2006,Sbisa2014}
for more details). Such a theory cannot describe a stable vacuum
and therefore has to be ruled out. The origin of the fast decay lies
in an integration over the entire phase space when computing scattering
amplitudes which diverge in the ultraviolet (UV) region. Therefore,
the only way to tame the ghost is to modify the integration in the
UV. In Ref. \cite{Carroll2003}, the authors suggested that new operators beyond the EFT would allow us to cut this integral and, therefore,
a theory with ghosts could theoretically be cured. In fact, it has been shown
that the vacuum in simple theories with two oscillators, of which one is a ghost, can indeed have a decay time that is larger than
the Hubble time \cite{Carroll2003,KaplanSundrum2005} (see also Refs. \cite{Dyda2012,Ramazanov2016} for discssions of ghosts in Chern-Simons and Ho\v{r}ava-Lifshitz theories, respectively). The energy
scale at which new physics might enter and break Lorentz invariance (LI) can be low enough
to slow down the vacuum decay sufficiently and circumvent any violation
of experimental constraints, but at the same time be high enough to be above the
cutoff of the EFT. In fact, a Lorentz breaking (LB) does not render the theory unappealing as long as it occurs above the EFT cutoff scale.

In this work, we discuss a theory of modified gravity that automatically
introduces a ghost instead of adding a simple ghost field by hand
to a well behaved theory. In fact, as will be shown, many properties of such a theory, like the decay
time of the vacuum, may be significantly different,
and therefore the conclusions from simpler toy models should not be adopted blindly. In order to modify GR suitably, we assume a massive graviton. Even though the
idea of studying massive gravity is very old \cite{FierzPauli1939},
ghost-free non-linear theories were discovered only recently \cite{HassanRosen2012c,Creminelli2005,deRhamGabadadze2010,HassanRosen2012b,HassanRosenSchmidt-May2012,deRhamGabadadzeTolley2011,HassanSchmidt-MayStrauss2012c}.
Here we use the so-called ghost-free de Rham-Gabadadze-Tolley (dRGT)
theory to construct a theory of a massive graviton with an
additional Boulware-Deser (BD) ghost \cite{BoulwareDeser1972}, which we then dub {\it haunted massive
gravity} (HMG). We first study the classical behavior of the theory for a Friedmann-Lema\^{i}tre-Robertson-Walker (FLRW) background
in order to identify the models that do not introduce
potentially dangerous instabilities already at this level. With HMG we find the first theory of a canonical non-linear massive gravity which possesses models that are free of any background pathologies, and allows for dynamical, even self-accelerating, FLRW solutions. We finally discuss the quantum stability of the viable models by computing the ghost and vacuum decay rates in HMG.

Although the theory that is discussed in this work has some nice features,
e.g., it can provide a solution to the dark energy problem, it is certainly not the most promising contender of GR. For example, the construction of the mass term is mainly based on keeping simplicity, and the strong coupling scale of the theory is very low, losing predictivity on small scales. This work, however, does not intend to present a new theory of modified gravity in order to address the problems with GR, but rather to examine the behavior of ghosts appearing in more realistic
theories of gravity. We find that the scattering processes that dominate in the ghost and vacuum decay rates
do not coincide with those that appear in theories with two coupled
canonical scalar fields where one of the fields is a ghost. While in such a
simple scenario the decay time has been found to scale only quadratically
with the cutoff at which Lorentz violation (LV) occurs \cite{Carroll2003}
(see also Refs. \cite{KaplanSundrum2005,Cline2003} for a discussion
on decay rates for other setups), we find a completely different scaling for HMG. Furthermore, we expect the
type of interaction that we find in this work as the most important one in HMG to
in fact determine the decay time also in many other theories of gravity
with a present ghost mode.

\section{HMG with $\text{d}$RGT limit}

Since we are interested in a theory of a massive graviton with an additional
BD ghost, we could simply study any non-linear theory that does not
coincide with the dRGT theory. However, we would like to keep a ghost-free
limit, and therefore, we start with dRGT and modify the tuning between
the coefficients of interaction terms.

The dRGT massive gravity can be written as \cite{HassanRosen2011a,deRhamGabadadzeTolley2011}
\begin{equation}
S_{\text{dRGT}}=M_{\text{P}}^{2}\int \td^{4}x\sqrt{-g}\left[R+2m^{2}U(\mathbb{K})\right],
\end{equation}
where $U(\mathbb{K})$ is the mass term, which depends on the eigenvalues
of $\mathbb{K}\equiv\sqrt{g^{-1}f}-\mathbb{1}\equiv\mathbb{X}-\mathbb{1}$,
and is equivalent to $\sum_{n=0}^{3}\beta_{n}U_{n}$ with
$U_{n}\equiv e_{n}\left(\mathbb{X}\right)$, $e_{n}$ being the elementary symmetric polynomials of the eigenvalues
of $\mathbb{X}$. Additionally, $f$ is a non-dynamical symmetric rank-2 tensor field, and $\beta_{n}$ depend
on the two free dimensionless parameters of the theory, $\alpha_{3}$
and $\alpha_{4}$ \cite{HassanRosen2011a}:
\begin{align}
\beta_{0} & =6-4\alpha_{3}+\alpha_{4},\\
\beta_{1} & =-3+3\alpha_{3}-\alpha_{4},\\
\beta_{2} & =1-2\alpha_{3}+\alpha_{4},\\
\beta_{3} & =\alpha_{3}-\alpha_{4}.
\end{align}
Let us first discuss the simplest model, the so-called minimal model
\cite{HassanRosen2011a}, and choose $\alpha_{3}$ and $\alpha_{4}$
such that we can switch off the highest order interactions, i.e. $\beta_{2}=\beta_{3}=0$,
and obtain
\begin{align}
\alpha_{3}=\alpha_{4}=1\quad\Leftrightarrow\quad\beta_{0}=3,\;\beta_{1}=-1.
\end{align}
The action in this case becomes
\begin{align}
S_{\text{min}} & =M_{\text{P}}^{2}\int \td^{4}x\sqrt{-g}\left[R+2m^{2}\left(3-\left[\mathbb{X}\right]\right)\right],
\end{align}
where $\mbox{\ensuremath{\left[\mathbb{X}\right]}}$ denotes the trace of $\mathbb{X}$.
If we change the prefactors of the mass term, we then
change either the CC or the graviton mass (or make it tachyonic). Thus, in
order to introduce a ghost, we should switch on higher order interactions.
One possibility would be to remove the CC (which would make the model very appealing especially
for cosmology) and allow for $\beta_{1}$
and $\beta_{2}$ to be non-vanishing. We then find $\beta_{0}=\beta_{3}=0$ together with
\begin{align}
\alpha_{3}=\alpha_{4}=2\quad\Leftrightarrow\quad\beta_{1}=1=-\beta_{2},
\end{align}
which results in the action 
\begin{align}
S & =M_{\text{P}}^{2}\int \td^{4}x\sqrt{-g}\left[R+2m^{2}\left(\beta_{1}\left[\mathbb{X}\right]+\frac{1}{2}\beta_{2}\left(\left[\mathbb{X}\right]^{2}-\left[\mathbb{X}^{2}\right]\right)\right)\right]\\
 & =M_{\text{P}}^{2}\int \td^{4}x\sqrt{-g}\left[R+2m^{2}\left(\left[\sqrt{g^{-1}f}\right]-\frac{1}{2}\left(\left[\sqrt{g^{-1}f}\right]^{2}-\left[g^{-1}f\right]\right)\right)\right].
\end{align}
Note that this choice, like all other combinations that satisfy $\alpha_{3}+\alpha_{4}>0$,
does not lead to a Higuchi ghost, at least around an FLRW background
for large $H^{2}$ \cite{Fasiello2012}. This is important because
we will use this theory as a ghost-free limit which should
ensure not only the absence of the BD ghost but also the presence of five healthy
graviton degrees of freedom.

We now modify the theory to introduce a ghost.
The simplest way would be to modify the prefactor in one of the interaction
terms which in the linear theory corresponds to a violation of the
Fierz-Pauli (FP) tuning. However, we do not expect this modification to enable dynamical
FLRW solutions for a flat reference metric since the combination of
the Bianchi identities and the conservation of energy-momentum tensor
will still be a constraint for the scale factor, as it has been shown for
dRGT in Ref. \cite{DAmico2011}. One way out could be to introduce a
metric-dependent (and, thus, lapse-dependent) prefactor like $\alpha \left[g^{-1}f\right]$, with $\alpha \in \mathbb R$,
and study the action
\begin{equation}
S=M_{\text{P}}^{2}\int \td^{4}x\sqrt{-g}\left[R+2m^{2}\left(\left[\sqrt{g^{-1}f}\right]+\frac{1}{2}\left(\alpha\left[g^{-1}f\right]-1\right)\left(\left[\sqrt{g^{-1}f}\right]^{2}-\left[g^{-1}f\right]\right)\right)\right].
\end{equation}
Although this theory would certainly allow for dynamical FLRW backgrounds,
we found only unviable solutions for which the scale factor
would become imaginary or the lapse would cross zero, indicating an instability.
Therefore, we consider a slightly more complicated theory in which
both interaction terms are modified, and dub this theory haunted massive
gravity (HMG):
\begin{equation}
S_{\text{HMG}}=M_{\text{P}}^{2}\int \td^{4}x\sqrt{-g}\left[R+2m^{2}\left(\left(1-\alpha_{1}\left(g,f\right)\right)\left[\sqrt{g^{-1}f}\right]-\frac{1}{2}\left(1-\alpha_{2}\left(g,f\right)\right)\left(\left[\sqrt{g^{-1}f}\right]^{2}-\left[g^{-1}f\right]\right)\right)\right],\label{eq:action_HMG}
\end{equation}
with
\begin{align}
\alpha_{1}\left(g,f\right) & \equiv\bar{\alpha}_{1}\mathbb{X}^{2}=\bar{\alpha}_{1}g^{\alpha\beta}f_{\beta\alpha},\\
\alpha_{2}\left(g,f\right) & \equiv\bar{\alpha}_{2}\mathbb{X}^{2}=\bar{\alpha}_{2}g^{\alpha\beta}f_{\beta\alpha},
\end{align}
where $\bar{\alpha}_{i}$ are two dimensionless parameters.

This theory has some interesting properties. Firstly, the limit $\bar{\alpha}_{i}\rightarrow0$
corresponds to the ghost-free dRGT theory, whereas any other values
should introduce a new ghost degree of freedom as it does not coincide with dRGT,
the unique non-linear ghost-free theory of a massive graviton. Secondly,
the additional dynamical factors will enable us to have dynamical FLRW solutions
by modifying the Bianchi constraint, and, finally, we expect the vacuum
to decay more slowly at late times since $\left[g^{-1}f\right]\propto a^{-2}$
for FLRW backgrounds.%
\footnote{This could have an interesting impact on the phenomenology at early
times and might lead to an enhanced creation of particles. The relevant
time period would, however, presumably lie above the cutoff scale of the
theory.%
}

\section{Background cosmology}

From now on, let us fix the reference metric to a flat Minkowski background, i.e.,
$f_{\mu\nu}=\eta_{\mu\nu}$. Since massive gravity with $f_{\mu\nu}=\eta_{\mu\nu}$
breaks diffeomorphism invariance, the lapse of $g_{\mu\nu}$ must not be chosen arbitrarily. For an FLRW background we therefore choose
\begin{equation}
\td s^{2}=-N_{g}^{2}\td t^{2}+a^{2}\td \vec{x}^{2},
\end{equation}
with $N_{g}$ and $a$ denoting the lapse and the scale factor, respectively, and $t$ being cosmic time.
Varying the action (\ref{eq:action_HMG}) with respect to $g_{\mu\nu}$
yields
\begin{align}
-2m^{2}M_{\text{P}}^{2}\delta\left(\sqrt{-g}U_{1}\left(X\right)\right) & =-\sqrt{-g}\beta_{1}m^{2}M_{\text{P}}^{2}g^{\mu\alpha}Y_{(1)\,\alpha}^{\nu}\left(\sqrt{g^{-1}f}\right)\delta g_{\mu\nu},\\
-2m^{2}M_{\text{P}}^{2}\delta\left(\sqrt{-g}U_{2}\left(X\right)\right) & =\sqrt{-g}\beta_{2}m^{2}M_{\text{P}}^{2}g^{\mu\alpha}Y_{(2)\,\alpha}^{\nu}\left(\sqrt{g^{-1}f}\right)\delta g_{\mu\nu},
\end{align}
with
\begin{align}
Y_{(1)}\left(\mathbb{X}\right) & \equiv \mathbb{X}-\mathbb{1}\left[\mathbb{X}\right],\\
Y_{(2)}\left(\mathbb{X}\right) & \equiv \mathbb{X}^{2}-\mathbb{X}\left[\mathbb{X}\right]+\frac{1}{2}\mathbb{1}\left(\left[\mathbb{X}\right]^{2}-\left[\mathbb{X}^{2}\right]\right).
\end{align}
Furthermore, we need the variation of $\alpha_{i}$:
\begin{align}
\delta\alpha_{i}\left(g\right) & =-\bar{\alpha}_{i}g^{\alpha\mu}g^{\nu\beta}\eta_{\beta\alpha}\delta g_{\mu\nu}.
\end{align}

With this, the variation of the mass term yields 
\begin{equation}
\sqrt{-g}M_{\text{P}}^{2}V^{\mu\nu}\equiv-m^{2}\sqrt{-g}M_{\text{P}}^{2}\left[\left(\alpha_{1}\left(g\right)-1\right)g^{\mu\alpha}Y_{(1)\,\alpha}^{\nu}+\left(\alpha_{2}\left(g\right)-1\right)g^{\mu\alpha}Y_{(2)\,\alpha}^{\nu}-2U_{1}\left(\mathbb{X}\right)\frac{\delta\alpha_{1}\left(g\right)}{\delta g_{\mu\nu}}-2U_{2}\left(\mathbb{X}\right)\frac{\delta\alpha_{2}\left(g\right)}{\delta g_{\mu\nu}}\right].
\end{equation}
Combining the Bianchi identities with the assumption of a conserved energy-momentum
tensor leads to the Bianchi constraint
\begin{equation}
\nabla_{\mu}V^{\mu\nu}=0,
\end{equation}
which implies
\begin{align}
\left(1+3a^{-2}N_{g}^{2}\right)\left[N_{g}'\left(a\left(a\bar{\alpha}_{1}+6\bar{\alpha}_{2}\right)+2N_{g}\left(2a\bar{\alpha}_{1}+\bar{\alpha}_{2}\right)\right)\right.\nonumber \\
\left.+HN_{g}\left(-6a\bar{\alpha}_{2}+N_{g}^{2}\left(4\bar{\alpha}_{1}-(a-2)a\right)-2N_{g}\left(a\bar{\alpha}_{1}+\bar{\alpha}_{2}\right)+9\bar{\alpha}_{1}a^{-1}N_{g}^{3}\right)\right] & =0.\label{eq:Bianchi_constraint}
\end{align}
Here, $H^2\equiv a'/a$ is the Hubble rate, and a prime denotes a derivative with respect to $t$. In the limit $\bar{\alpha}_{i}\rightarrow0$, Eq. (\ref{eq:Bianchi_constraint}) fixes the scale factor confirming the no-go theorem for
FLRW solutions in dRGT massive gravity. In our case, however, we are
able to switch on the dynamics since the Bianchi constraint now depends on both the scale factor and the lapse. This constraint together with the Friedmann equation
\begin{align}
3H^{2} & =\rho+V_{00}\\
 & =\rho+\frac{m^{2}}{a^{4}N_{g}}\left[a^{3}\left(-\left(a\bar{\alpha}_{1}+6\bar{\alpha}_{2}\right)\right)-3a^{2}N_{g}\left(2a\bar{\alpha}_{1}+\bar{\alpha}_{2}\right)+3N_{g}^{3}\left(a\left((a-1)a-3\bar{\alpha}_{1}\right)+3\bar{\alpha}_{2}\right)\right]
\end{align}
and assuming a universe filled with dark matter only,
\begin{equation}
\rho\equiv\rho_{0}a^{-3},
\end{equation}
can be solved numerically. In the limit $a\ll 1$, the combination of the Bianchi constraint and the Friedmann equation provides
\begin{equation}
N_g = \pm \frac{1}{3}\sqrt{\frac{\bar \alpha_2}{\bar \alpha_1} a},
\end{equation}
and implies $H^2 \propto a^{-3}$. Therefore, we find a singularity for $a\rightarrow 0$. The time at which it occurs will be denoted by $t_c$, i.e., $N_g (t_c) = 0$.

Interestingly, for a given $\bar{\alpha}_{1}$ one can find a value for the parameter $\bar{\alpha}_{2}$ that maximizes the timescale of the background evolution by reaching $t_c \rightarrow 0$. One example
for such a model is
\begin{equation}
\bar{\alpha}_{1}=0.9\;\Rightarrow\;\bar{\alpha}_{2}\simeq0.1025.\label{eq:model4plot}
\end{equation}

By solving the background equations numerically, we can search for the
parameter region that leads to $t_c = 0$ and find
that it can be fitted very well linearly with
\begin{equation}
\bar{\alpha}_{2}\simeq\frac{1}{6}\bar{\alpha}_{1}-\frac{2}{45}.\label{eq:stability_curve}
\end{equation}

If we promote the maximization of the classical timescale to a constraint,
then this model will effectively lose one free parameter. Furthermore, Eq. (\ref{eq:stability_curve}) indicates that both $\bar \alpha_1$ and $\bar \alpha_2$ can be of $\mathcal O \left(1\right)$.

\begin{figure}[t]
\includegraphics[width=0.45\textwidth]{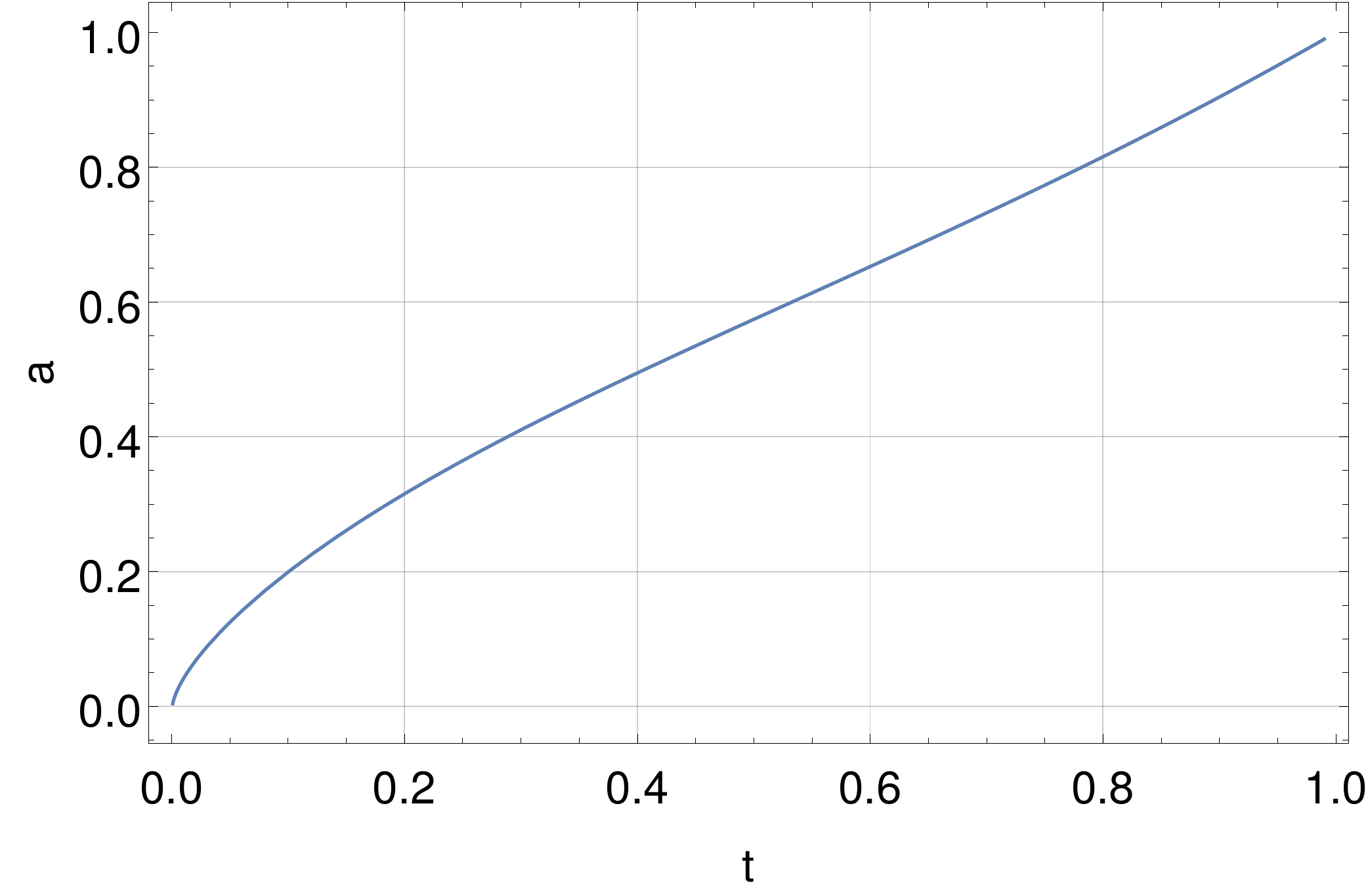}$\quad$\includegraphics[width=0.45\textwidth]{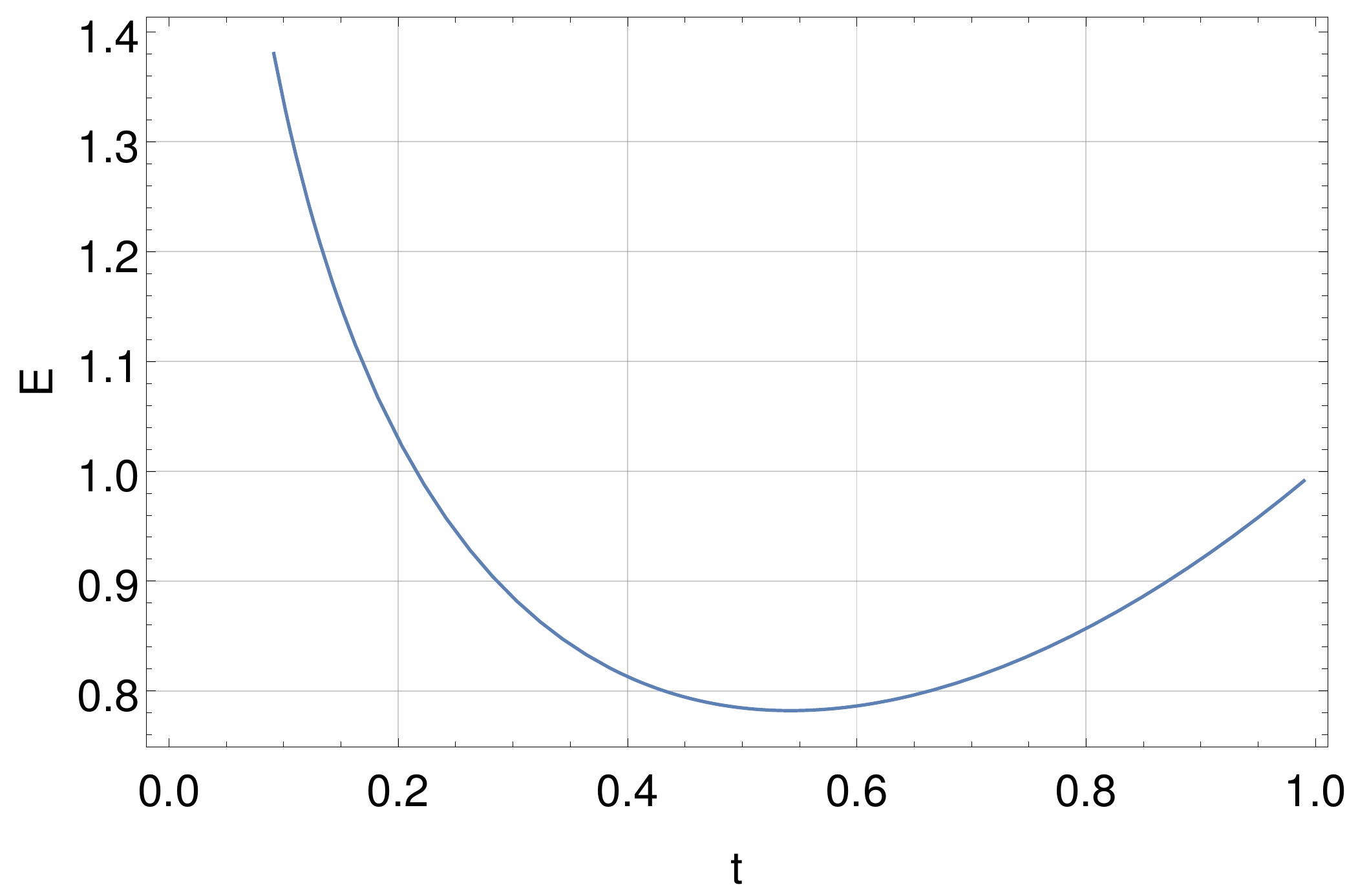}\\
\includegraphics[width=0.45\textwidth]{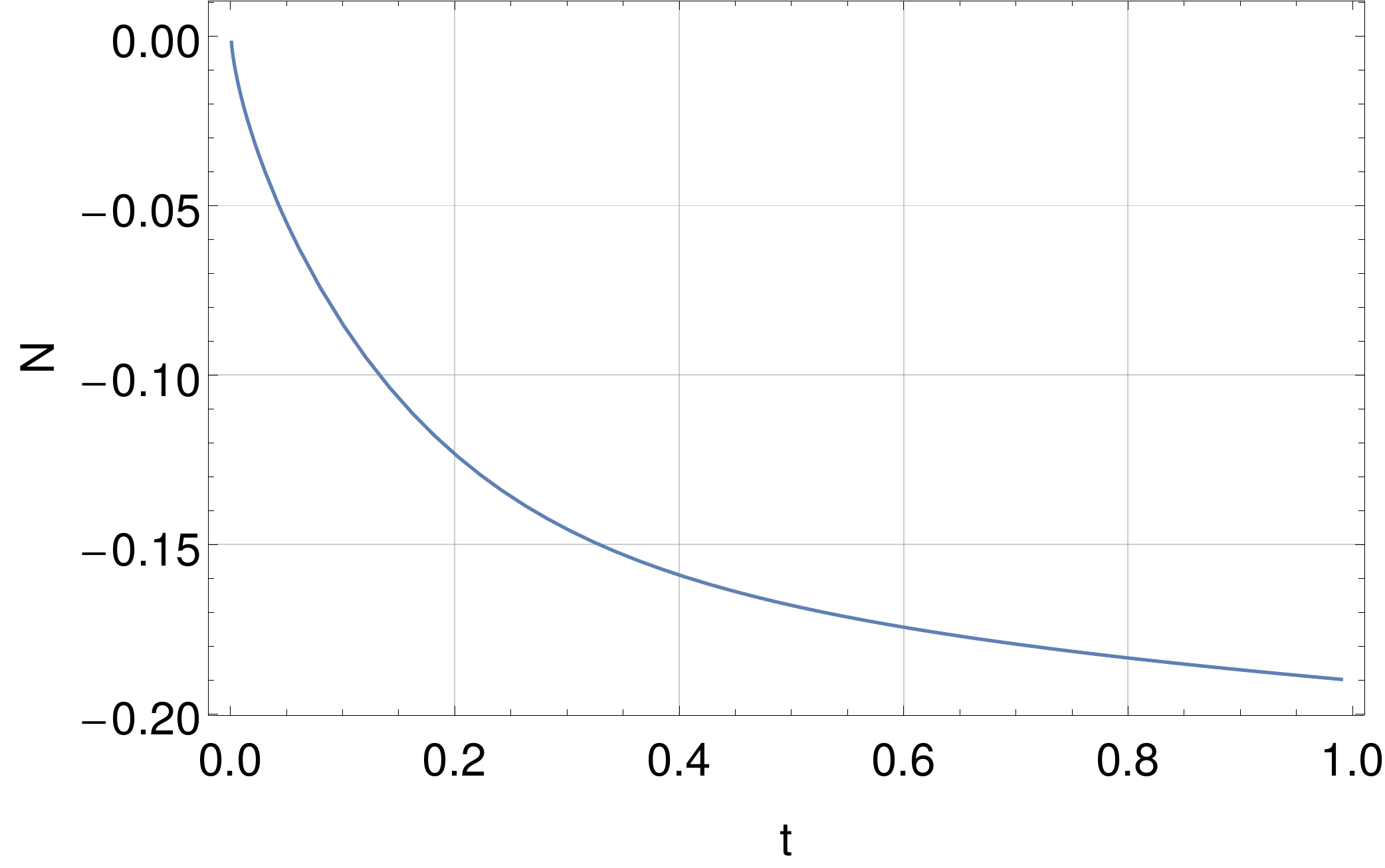}$\quad$\includegraphics[width=0.45\textwidth]{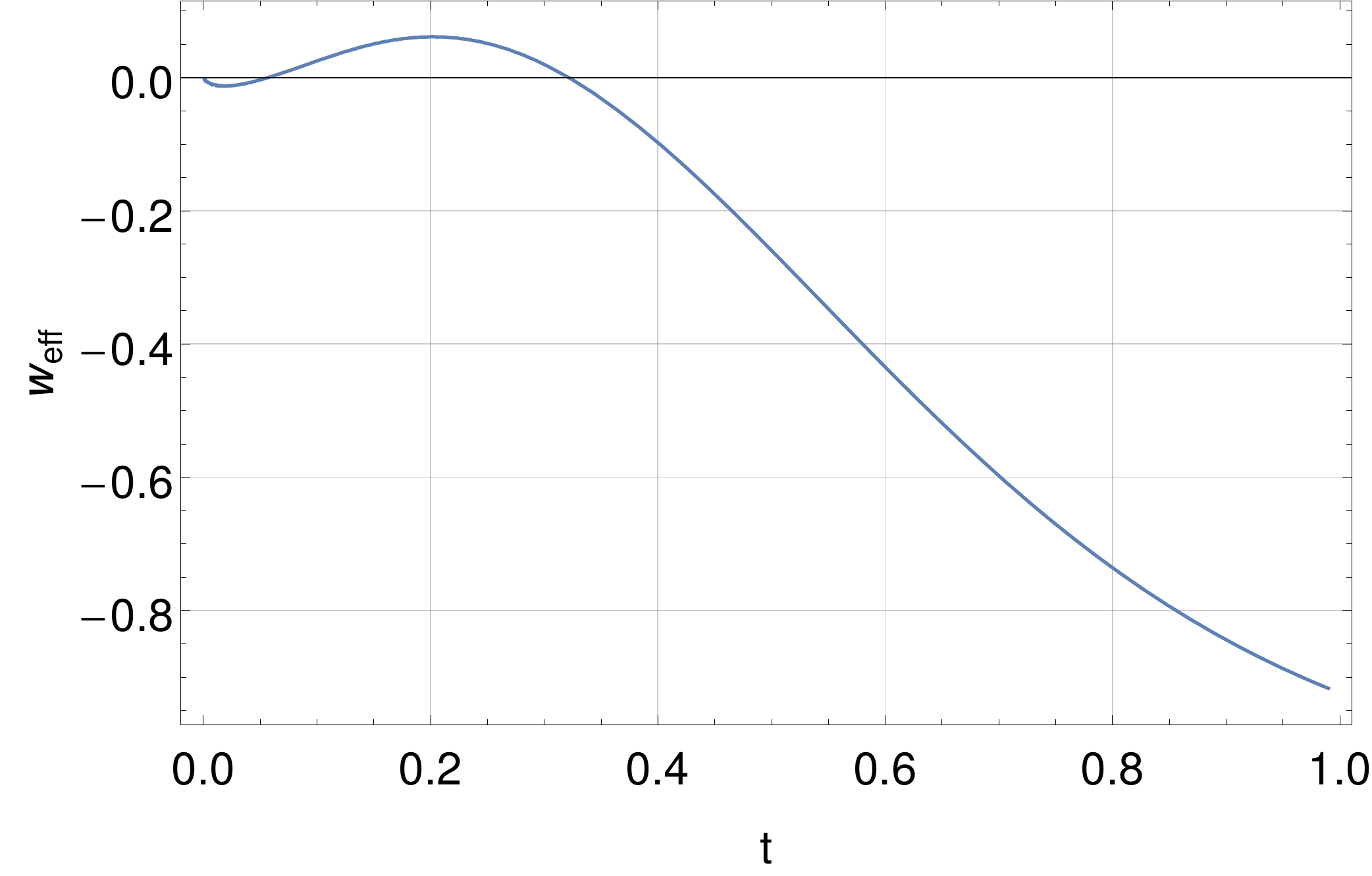}

\protect\caption{Numerical solution of the FLRW background evolution in HMG, corresponding to the model with
$\bar{\alpha}_{1}=0.9$ and $\bar{\alpha}_{2}\simeq0.1025$. The plots show the scale factor (upper left), the expansion rate
(upper right), the lapse (lower left), and the effective equation of state
(lower right). All the quantities are plotted versus the cosmic time, $t$, scaled in such a way that $a(t=1)=1$.}
\label{fig:background_numerical_solution}

\end{figure}

Surprisingly, after solving the background equations for the model (\ref{eq:model4plot}) numerically, we find that at late times the effective equation of state parameter
$w_{\text{eff}}<-1/3$ indicates a period of self-acceleration and we thus
have found a candidate for a model that could be able to provide a
solution to the dark energy problem. See Fig. \ref{fig:background_numerical_solution}
for a numerical solution of the background evolution, corresponding to model (\ref{eq:model4plot}).

\section{Second order action for HMG around Minkowski}

In order to compute the vacuum decay, we have to first identify the ghost mode.
For this, we perturb the background, expand the action to second order,
and finally integrate out \footnote{Since we are interested in computing only tree-level diagrams in order to discuss the quantum behavior of the theory, the elimination of all auxiliary fields in the action by using their EoM is equivalent to properly integrating out these fields.} all auxiliary fields.

\subsection{Gravity sector}

We now choose to work with perturbations around a Minkowski background. Note that we expect a generalization to an FLRW background to merely modify the decay rate insignificantly. In fact, corrections from an FLRW background are proportional to $H/k$ and become negligible in the high-momentum limit, on which we will focus later. Furthermore, ignoring the cosmological
expansion and, therefore, a smaller volume at early times, should then correspond
to maximizing the decay rate and computing an upper bound for a more
realistic scenario.

In this case, the Bianchi constraint (\ref{eq:Bianchi_constraint})
enforces the lapse to be a constant; here we set $N_{g}=1$.
Furthermore, in order to excite the scalar BD ghost we consider the following scalar perturbations $\delta g$ around the background $\bar g\equiv \eta$:
\begin{equation}
\td s_{\delta g}^{2}=2\left[-\Psi \td t^{2}+B_{,i}\td x^{i}\td t+\left(\Phi\delta_{ij}+E_{,ij}\right)\td x^{i}\td x^{j}\right].\label{eq:pert_ansatz}
\end{equation}
The Einstein-Hilbert action at second order therefore reads
\begin{equation}
S_{\text{EH}}^{(2)}=4M_{\text{P}}^{2}\int \td^{4}x\left(\Phi_{i}^{2}+2\Phi_{i}\Psi_{i}-2\Phi'\Delta E'-3\Phi'^{2}-2B_{i}\Phi_{i}'\right),\label{eq:action_EH}
\end{equation}
where we have used the notation
\begin{equation}
X_{i}Y_{i}\equiv\sum_{j}X_{,j}Y_{,j}\;,\qquad\Delta X\equiv\sum_{j}X_{,jj}.
\end{equation}

We now expand the mass term of action (\ref{eq:action_HMG}) to second order and use
\begin{align}
\sqrt{g^{-1}\eta} & \simeq\sqrt{\bar{g}^{-1}\eta}\left[1-\frac{1}{2}\bar{g}^{-1}\left(\delta g\right)+\frac{3}{8}\bar{g}^{-1}\left(\delta g\right)\bar{g}^{-1}\left(\delta g\right)\right]\\
 & =\mathbb{1}-\frac{1}{2}\eta\left(\delta g\right)+\frac{3}{8}\eta\left(\delta g\right)\eta\left(\delta g\right),
\end{align}
to obtain
\begin{align}
\left(1-\bar{\alpha}_{1}\left[g^{-1}\eta\right]\right)\left[\sqrt{g^{-1}\eta}\right]\simeq & \left[\mathbb{1}-\frac{1}{2}\eta\left(\delta g\right)+\frac{3}{8}\eta\left(\delta g\right)\eta\left(\delta g\right)\right]\nonumber\\
& -\bar{\alpha}_{1}\left(\left[\mathbb{1}\right]-\left[\left(\delta g\right)\eta\right]+\left[\eta\left(\delta g\right)\eta\left(\delta g\right)\right]\right)\left(\left[\mathbb{1}\right]-\left[\frac{1}{2}\eta\left(\delta g\right)\right]+\left[\frac{3}{8}\eta\left(\delta g\right)\eta\left(\delta g\right)\right]\right)\\
\simeq & 4\left(1-4\bar{\alpha}_{1}\right)-\left(\frac{1}{2}-6\bar{\alpha}_{1}\right)\left[\eta\left(\delta g\right)\right]-\frac{1}{2}\bar{\alpha}_{1}\left[\eta\left(\delta g\right)\right]^{2}+\left(\frac{3}{8}-\frac{11}{2}\bar{\alpha}_{1}\right)\left[\eta\left(\delta g\right)\eta\left(\delta g\right)\right]
\end{align}
and
\begin{align}
\left(1-\bar{\alpha}_{2}\left[g^{-1}\eta\right]\right)\left(\left[\sqrt{g^{-1}\eta}\right]^{2}-\left[g^{-1}\eta\right]\right)\simeq & \left(1-\bar\alpha_2\left(4-\left[\left(\delta g\right)\eta\right]+\left[\eta\left(\delta g\right)\eta\left(\delta g\right)\right]\right)\right) \times \nonumber\\
& \times \left(\left[\mathbb{1}-\frac{1}{2}\eta\left(\delta g\right)+\frac{3}{8}\eta\left(\delta g\right)\eta\left(\delta g\right)\right]^2-4+\left[\left(\delta g\right)\eta\right] - \left[ \eta \left(\delta g\right) \eta \left(\delta g\right) \right] \right)\\
\simeq & 12\left(1-4\bar{\alpha}_{2}\right)-3\left(1-8 \bar{\alpha}_{2}\right)\left[\left(\delta g\right)\eta\right]+\left(\frac{1}{4}-4\bar{\alpha}_{2}\right)\left[\eta\left(\delta g\right)\right]^{2}\nonumber\\
&+2\left(1-10\bar{\alpha}_{2}\right)\left[\eta\left(\delta g\right)\eta\left(\delta g\right) \right].
\end{align}
Finally, with 
\begin{align}
\sqrt{-g} & \simeq\frac{1}{4}\sqrt{-\bar{g}}\left(4+2\left(\delta g\right)_{\;\mu}^{\mu}-\left(\delta g\right)_{\mu\nu}\left(\delta g\right)^{\mu\nu}+\frac{1}{2}\left(\left(\delta g\right)_{\;\mu}^{\mu}\right)^{2}\right)\\
 & =\frac{1}{4}\left(4+2\left[\eta\left(\delta g\right)\right]-\left[\eta\left(\delta g\right)\eta\left(\delta g\right)\right]+\frac{1}{2}\left[\eta\left(\delta g\right)\right]^{2}\right)
\end{align}
we find the second order action of the mass term as
\begin{align}
S_{\text{mass}}^{(2)}=M_{\text{P}}^{2}m^{2}\int \td ^{4}x & \left(\frac{1}{4}+\bar\alpha_1-2\bar\alpha_2\right)\left[\eta\left(\delta g\right)\right]^{2}-\left(\frac{1}{4}+3\bar{\alpha}_{1}-8\bar{\alpha}_{2}\right)\left[\eta\left(\delta g\right)\eta\left(\delta g\right)\right].\label{eq:pot_2nd}
\end{align}
For the ansatz (\ref{eq:pert_ansatz}), this becomes
\begin{align}
S_{\text{mass}}^{(2)}= & \frac{1}{2} M_{\text{P}}^{2}m^{2}\int \td ^{4}x\left[-c_{1}B\Delta B+c_{2}\left(\Psi^{2}+\left(\Delta E\right)^{2}\right)+8c_{3}\Phi\Delta E+12c_{3}\Phi^{2}+4c_{4}\Psi\left(\Delta E+3\Phi\right)\right],\label{eq:action_pot}
\end{align}
where we have defined the parameters $c_{i}$ as
\begin{align}
c_{1} & \equiv1+12\bar{\alpha}_{1}-32\bar{\alpha}_{2},\label{eq:def_c1}\\
c_{2} & \equiv-16\bar{\alpha}_{1}+12\bar{\alpha}_{2},\label{eq:def_c2}\\
c_{3} & \equiv1+4\bar{\alpha}_{2},\label{eq:def_c3}\\
c_{4} & \equiv1+4\bar{\alpha}_{1}-8\bar{\alpha}_{2}\label{eq:def_c4}.
\end{align}
Therefore, the (minimal) ghost-free massive gravity corresponds to
the limit $c_1,c_3,c_4\rightarrow1$ and $c_2\rightarrow 0$. Note that we should not necessarily
expect a smooth limit due to the change in the number of degrees of freedom in HMG compared to dRGT.

\subsection{Full action including matter}

For simplicity, we assume the matter sector to contain only a single, minimally-coupled, scalar field
$\varphi$ with mass $m_{\varphi}$, i.e.,
\begin{equation}
S_{\text{matter}}=-\int \td ^{4}x\sqrt{-g}\left(\partial^{\mu}\varphi\partial_{\mu}\varphi+m_{\varphi}^{2}\varphi^{2}\right),\label{eq:action_matter}
\end{equation}
and consider the weak-field limit where only the gravitational sector is expanded to second order and matter is kept unperturbed. The kinetic and mass terms of $\varphi$ are
then given by its coupling to the background metric.

By combining the actions (\ref{eq:action_EH}), (\ref{eq:action_pot}), and (\ref{eq:action_matter}),
we obtain the final leading order action of HMG, containing
a Boulware-Deser ghost and a matter field, around Minkowski (modulo total
derivatives),
\begin{align}
S_{\text{HMG}}^{(2)}=\int \td ^{4}x & \left[4M_{\text{P}}^{2}\left(\Phi_{i}^{2}-2\Delta\Phi\Psi-2\Phi'\Delta E'-3\Phi'^{2}-2B_{i}\Phi_{i}'\right)\right.\nonumber \\
 & \left.+\frac{1}{2}m^{2}M_{\text{P}}^{2}\left(c_{1}B_{i}^{2}+c_{2}\left(\Psi^{2}+\left(\Delta E\right)^{2}\right)+8c_{3}\Phi\Delta E+12c_{3}\Phi^{2}+4c_{4}\Psi\left(\Delta E+3\Phi\right)\right)\right.\nonumber \\
 & \left.-\left(1+B_{i}^{2}-\left(\Delta E\right)^{2}+3\Phi^{2}+6\Phi\Psi-\Psi^{2}+2\Delta E\left(\Phi+\Psi\right)\right)X_{\varphi}\right],\label{eq:action_HMG_2nd_order}
\end{align}
where we have defined
\begin{equation}
X_{\varphi}\equiv-\varphi'^{2}+\varphi_{i}^{2}+m_{\varphi}^{2}\varphi^{2}.\label{eq:def_X_varphi}
\end{equation}

In total, all the scalar potentials $\Phi$, $\Psi$, $B$, $E$, and the matter field
$\varphi$ should describe at most three propagating scalar degrees of freedom: one
helicity-0 mode of the graviton, one scalar field from the matter sector,
and one additional Boulware-Deser ghost. All these scalar degrees of freedom are,
however, not always excited around all backgrounds. Since we are interested
in the interaction of the ghost with the matter field, we need to
ensure that the BD ghost is indeed a propagating mode in Eq. (\ref{eq:action_HMG_2nd_order}).
To see this, we first integrate out all the auxiliary fields by using
their EoM
\begin{equation}
\frac{\partial\mathcal{L}}{\partial X}-\partial_{t}\left(\frac{\partial\mathcal{L}}{\partial X'}\right)-\partial_{i}\left(\frac{\partial\mathcal{L}}{\partial X_{i}}\right)+\partial_{t}^{2}\left(\frac{\partial\mathcal{L}}{\partial X''}\right)+\partial_{i}^{2}\left(\frac{\partial\mathcal{L}}{\partial\left(\partial_{i}^{2}X\right)}\right)=0.
\end{equation}
For $X\in\left\{ \Psi,\, B,\,\Delta E\right\} $ this leads to
\begin{align}
\Psi & =\frac{8\Mp^2\Delta\Phi-\left(\Delta E+3\Phi\right)\left(2c_{4}m^{2}M_{\text{P}}^{2}+X_{\varphi}\right)}{c_{2}m^{2}M_{\text{P}}^{2}-X_{\varphi}},\label{eq:EoM_auxvars_Psi}\\
B_{i} & =\frac{8\Mp^2\Phi_{i}'}{c_{1}m^{2}M_{\text{P}}^{2}+X_{\varphi}},\label{eq:EoM_auxvars_Bi}\\
\Delta E & =-\frac{\Phi\left(4c_{3}m^{2}M_{\text{P}}^{2}+X_{\varphi}\right)+\Psi\left(2c_{4}m^{2}M_{\text{P}}^{2}+X_{\varphi}\right) + 8 \Mp^2 \Phi''}{c_{2}m^{2}M_{\text{P}}^{2}-X_{\varphi}}.\label{eq:EoM_auxvars_DelE}
\end{align}
Solving this set of equations for $\Psi$, $B$, and $\Delta E$ as functions of $\Phi$, $X_\varphi$, and their derivatives, yields
\begin{align}
\Psi & =\frac{8\Mp^2 \Delta\Phi\left(c_{2}m^{2}M_{\text{P}}^2-X_{\varphi}\right)-\left(2 c_4 m^2 \Mp^2 + X_{\varphi}\right)\left[ \Phi \left( \left(3c_2-4c_3\right)m^2 \Mp^2 - 4X_\varphi \right)-8 \Mp^2 \Phi'' \right]}{\left(c_{2}^2-4c_{4}^2\right)m^{4}\Mp^4-2\left(c_{2}+2c_{4}\right)m^{2}M_{\text{P}}^{2}X_{\varphi}},\label{eq:sol_Psi}\\
B_{i} & =\frac{8\Mp^2\Phi_{i}'}{c_{1}m^{2}M_{\text{P}}^{2}+X_{\varphi}},\label{eq:sol_B}\\
\Delta E & =\frac{8\Delta\Phi\left(2c_{4}m^{2}M_{\text{P}}^{2}+X_{\varphi}\right) +\Phi\left[4\left(c_{2}c_{3}-3c_{4}^{2}\right)m^{4}M_{\text{P}}^{4} +\left(c_{2}-4c_{3}-12c_{4}\right)m^{2}M_{\text{P}}^{2}X_{\varphi}-4X_{\varphi}^{2}\right] +8\Mp^2 \Phi'' \left( c_2 m^2 \Mp^2 - X_\varphi \right)}{-\left(c_{2}^2-4c_{4}^2\right)m^{4}\Mp^4+2\left(c_{2}+2c_{4}\right)m^{2}M_{\text{P}}^{2}X_{\varphi}}.\label{eq:sol_DeltaE}
\end{align}

Note that the determinant of the mixing matrix for $\Psi$ and $\Delta E$ in Eqs. (\ref{eq:EoM_auxvars_Psi}) and (\ref{eq:EoM_auxvars_DelE}) is proportional to the factors $m^2$ and $c_2+2c_4=2-8\bar\alpha_1 +32\bar\alpha_2$. If one of these terms vanishes, the mixing matrix becomes singular, which is equivalent to it having a zero eigenvalue. These eigenvalues correspond to the kinetic terms of the diagonalized degrees of freedom (given by the eigenvectors). Therefore, the singular situation corresponds to a combination of the auxiliary fields losing its kinetic term, and leads to a strong coupling and a breakdown of perturbativity.

Finally, the full second order HMG action
\begin{equation}
S_{\text{HMG}}^{(2)}=S_{\text{EH}}^{(2)}+S_{\text{mass}}^{(2)}+S_{\text{matter}}
\end{equation}
can be written as
\begin{equation}\label{action_HMG_ghost_matter}
S_{\text{HMG}}^{(2)}\left[\Phi,\varphi\right]=S_{\text{m}}^{(2)}\left[\Phi,\varphi\right]+S_{\text{kin}}^{(2)}\left[\Phi,\varphi\right]+S_{\text{int}}^{(2)}\left[\Phi,\varphi\right].
\end{equation}
The action now depends only on two remaining interacting massive scalar fields $\varphi$ and $\Phi$
described by the mass term
\begin{equation}
S_{\text{m}}^{(2)}=-\int \td ^{4}x\left(-6c_{3}m^{2}\Mp^2 \Phi^{2}+m_{\varphi}^{2}\varphi^{2}\right),
\end{equation}
and rather complicated kinetic and interaction terms, $S_{\text{kin}}^{(2)}$ and $S_{\text{int}}^{(2)}$, respectively. Because the action (\ref{eq:action_HMG_2nd_order}) contains a term that is proportional to $\Phi' \Delta E'$, we find, after integrating out $\Delta E$, terms that include $\left(\Phi''\right)^2$.  The occurrence of fourth-order derivatives in $\Phi$ signals that the theory is inevitably plagued by an Ostrogradsky ghost. In total, we expect a composition of three scalar degrees of freedom consisting of a helicity-0 mode from the graviton, a matter field, and a ghost. In order to analyze the interactions between the ghost and the other degrees of freedom, we need to decouple all of them.

\subsection{Decoupling of the ghost}

\subsubsection{Decoupling in vacuum}

Before analyzing the UV limit, i.e., $X_\varphi \gg m^2 \Mp^2$, we study the simpler case first in which the matter field is absent, i.e., $X_\varphi = 0$. The action can then be written as

\begin{equation}
S_{\text{HMG}}^{(2)}=\int \td^4 x \left[ C_1 \left(\Delta \Phi\right)^2 +C_2 \Phi \Delta \Phi + C_3 \Phi'' \Delta \Phi + C_4 \Phi \Phi'' + C_5 \left(\Phi''\right)^2 + C_6 \Phi^2 \right],\label{eq:action_Phi}
\end{equation}
where
\begin{align}
C_1 &\equiv -\frac{32 c_2 \Mp^2}{m^2 \left(c_2^2-4 c_4^2\right)}, &\quad C_2 &\equiv -\frac{4 \Mp^2 \left(c_2^2-12 c_2 c_4+4 c_4 (4 c_3-c_4)\right)}{c_2^2-4 c_4^2},\\
C_3 &\equiv -\frac{32 \Mp^2 \left(4 c_4 (c_1-c_4)+c_2^2\right)}{c_1 m^2 \left(c_2^2-4 c_4^2\right)}, &\quad C_4 &\equiv \frac{4 \Mp^2 \left(3 c_2^2-8 c_2 c_3+12 c_4^2\right)}{c_2^2-4 c_4^2},\\
C_5 &\equiv -\frac{32 c_2 \Mp^2}{m^2 \left(c_2^2-4 c_4^2\right)}, &\quad C_6 &\equiv \frac{2 m^2 \Mp^2 (3 c_2-4 c_3) \left(c_2 c_3-3 c_4^2\right)}{c_2^2-4 c_4^2}.
\end{align}

In order to make the additional scalar degree of freedom manifest we can try to find an equivalent action that descibes two fields with at most second derivatives instead of one field having fourth order derivatives. A special case where the interaction term is just $\left(\square \Phi\right)^2$ has already been presented in Ref.~\cite{Creminelli2005}.

For this, we introduce an auxiliary field $\chi$ together with seven unknown constants $D_i$, and consider a general action of two scalars $\Phi$ and $\chi$ that contains at most second-order derivatives,

\begin{equation}\label{eq:action_Phi_chi}
S'=\int \td^4 x \left[ D_1 \Phi \Phi'' + D_2 \Phi \Delta \Phi + D_3 \chi \Phi'' + D_4 \chi \Delta \Phi + D_5 \chi^2 + D_6 \chi \Phi + D_7 \Phi^2\right].
\end{equation}

The coupling to $\chi$ is constructed such that this auxiliary field can easily be integrated out by using its equation of motion,
\begin{equation}
\chi = -\frac{1}{2 D_5}\left( D_6 \Phi + D_3 \Phi'' + D_4 \Delta \Phi \right).
\end{equation}

We would then obtain an action that looks similar to the action (\ref{eq:action_Phi}) with which we started, except for the coefficients that will now depend on the constants $D_i$. Since we are interested in finding an equivalent action with two degrees of freedom, we equate these coefficients and solve for the unknown constants $D_i$. Interestingly, a solution does exist only if
\begin{equation}
C_1 =\frac{C_3^2}{4C_5}\;\;\Leftrightarrow\;\;  C_1 \left(\Delta \Phi\right)^2 + C_3 \Phi'' \Delta \Phi + C_5 \left(\Phi''\right)^2 = \left(\sqrt C_1 \Delta \Phi + \sqrt C_5 \Phi''\right)^2,
\end{equation}
which, as one can easily check, is also satisfied for HMG.\footnote{This condition enforces the theory to be covariant. Since we have started with a covariant theory and then performed a time-space splitting, it is indeed expected that this condition is satisfied. For theories that violate this constraint due to terms that break covariance, this does, however, not imply that there is no ghost but rather that our ansatz is not sufficient. This could indicate that the theory does not propagate only one but more degrees of freedom.} After fixing the redundancy due to a free rescaling of the actions by choosing $D_5 = - m^2 \Mp^2$, \footnote{This choice does also ensure real coefficients for parameter values that we will focus on later. Otherwise, if $c_2^2 - 4 c_4^2 $ is negative, the coefficient $D_5$ should be positive such that $\sqrt C_5$ is real.} we find

\begin{align}
D_1 &=C_4 \pm \sqrt C_5 D_6, \;\; & D_2 &= C_2 \pm \frac{C_3 D_6}{2\sqrt C_5}, \;\;& D_3 &=\mp 2 \sqrt C_5,\label{eq:sol_Di_1}\\
D_4 &= \mp \frac{C_3}{\sqrt C_5}, \;\; & D_7 &=C_6 - \frac{1}{4}D_6^2.\label{eq:sol_Di_2}
\end{align}

We can now introduce two scalar fields $\pi$ and $\ghost$ (to be pronounced ``phi spectre''), described by a superposition of $\Phi$ and $\chi$, which can finally be decoupled with the transformations

\begin{align}
\Phi \longrightarrow A_1 \pi - A_2 \ghost\quad\text{and}\quad\chi \longrightarrow \ghost,
\end{align}

where $A_1$ and $A_2$ are free coefficients. They can be used to diagonalize the mass terms,
\begin{equation}\label{eq:action_matter_decoupled_vacuum}
S'_m=\int \td^4 x \left[ D_5 \ghost^2 + D_6 \ghost \left(A_1 \pi  -A_2 \ghost\right) + D_7 \left( A_1 \pi - A_2 \ghost \right)^2  \right],
\end{equation}
 to obtain the physical degrees of freedom which can be achieved by setting
\begin{align}
A_1 D_6 = 2 A_1 A_2 D_7.
\end{align}

For the choice $A_1 = 1$ and $D_6 = m^2 \Mp^2$, and using the solution corresponding to the upper signs in Eqs. (\ref{eq:sol_Di_1}) and (\ref{eq:sol_Di_2}), we see that if $c_2^2 - 4 c_4^2 > 0$ (the parameter region that we are interested in) then the prefactor in the kinetic terms for $\ghost$ is negative and, therefore, describes a (BD) ghost, whereas the one for $\pi$ is positive and, thus, corresponds to the healthy helicity-0 mode. From Eq. (\ref{eq:action_matter_decoupled_vacuum}) we can read off the diagonalized mass terms for both scalars and find
\begin{align}
m^2_{\ghost} \Mp^2 &= \frac{4D_5 D_7 - D_6 ^2}{4 D_7} = \frac{-4m^2 \Mp^2 D_7 - m^4 \Mp^4}{4 D_7},\\
m^2_{\pi} \Mp^2 &= D_7.
\end{align}
Since $D_7$ is positive if Eq. (\ref{eq:stability_curve}) is satisfied, this indicates that the ghost is indeed a tachyon.

\subsubsection{Decoupling in the presence of matter}

So far, we have been able to decouple the ghost and the helicity-0 in the absence of an additional matter field. Due to integrating out all auxiliary fields in the full action (\ref{eq:action_HMG_2nd_order}), the coupling between matter and both the ghost and the helicity-0 mode is, however, not trivial and requires a proper decoupling of all present degrees of freedom. Fortunately, the procedure is conceptually similar to what has been done in the vacuum case. Furthermore, we can simplify the calculations by considering the small scale limit $X_\varphi \gg m^2 \Mp^2$. Because the analysis nevertheless becomes a bit lenghtier, we present some itermediate steps in Appendix \ref{app_dec_three_dof}.

In the presence of a matter field $\varphi$ and assuming small scales, the action can be decomposed as

\begin{align}\label{eq:action_Phi_phi}
S_{\text{HMG}}^{(2)}=\int \td^4 x &\left[ \left(\Phi''\right)^2 \left(C_1 \Phi^2 \varphi^2 + C_2 \Phi \varphi + C_3 \right) + \left(\varphi''\right)^2 \left(C_4 \Phi^2 \varphi^2 + C_5 \Phi \varphi + C_6 \right) \right.\\
&\left. + \left(\Delta \Phi\right)^2 \left(C_7 \Phi^2 \varphi^2 + C_8 \Phi \varphi + C_9 \right) + \left(\Delta \varphi\right)^2 \left(C_{10} \Phi^2 \varphi^2 + C_{11} \Phi \varphi + C_{12} \right) +C_{13}\Delta \Phi \Delta \varphi \Phi \varphi  \right.\nonumber\\ 
&\left.  + \Phi'' \left(C_{14}\Delta \varphi \Phi \varphi+ C_{15}\varphi'' \Phi \varphi + C_{16}\Delta \Phi \varphi^2 \Phi^2 +C_{17} \Delta \Phi \varphi \Phi +C_{18} \Delta \Phi + C_{19}\Delta \varphi \varphi^2 \Phi^2 +C_{20} \Delta \Phi \varphi \Phi^2\right) \right.\nonumber\\ 
&\left. +\Phi'' \left( C_{21}\Delta \varphi \varphi^2 \Phi^2 + C_{22}\Delta \varphi + C_{23}\varphi'' \varphi^2 \Phi^2 + C_{24}\varphi'' \right)   \right.\nonumber\\ 
&\left.+ \Delta \Phi \left( C_{25}\varphi'' \varphi^2 \Phi^2 + C_{26}\varphi'' + C_{27}\Delta \varphi \varphi^2 \Phi^2 + C_{28}\Delta \varphi \right) + C_{29}\Delta \varphi \varphi'' \Phi \varphi + C_{30}\Delta \varphi \varphi'' \right.\nonumber\\ 
&\left. +\Phi''\left( C_{31}\Phi \varphi^2 +C_{32} \Phi \right) + \varphi'' \left( C_{33}\Phi^2 \varphi^3 + C_{34}\Phi^2 \varphi + C_{35}\varphi \right) + \varphi \Phi \left( C_{36}\Phi \varphi^2 + C_{37}\Phi \right)    \right.\nonumber\\ 
&\left. + \Delta \varphi \left( C_{38}\Phi^2 \varphi^3 +C_{39} \Phi^2 \varphi +C_{40} \varphi \right) + C_{41}\Phi^2 \varphi^4 + C_{42}\Phi^2 \varphi^2 +C_{43} \varphi^2 + C_{44}\Phi^2  \right].\nonumber
\end{align}

Note that for HMG, some of the constants $C_i$ do indeed vanish. All of them are explicitly listed in Eq. (\ref{coeff_Ci_Phi_phi}). Again, we start with an ansatz for an action that explicitly describes three degrees of freedom with at most second-order derivatives,
\begin{align}\label{eq:action_Phi_phi_chi}
S_{\text{HMG}}'^{(2)}=\int \td^4 x &\left[  \Phi'' \left( D_{1}\Phi \varphi^2 + D_{2}\Phi \right) + \varphi'' \left( D_{3}\varphi^3 \Phi^2 + D_{4}\varphi \Phi^2 + D_{5}\varphi \right) \right.\\
&\left. + \Delta \Phi \left( D_{6}\Phi \varphi^2 + D_{7}\Phi \right) + \Delta \varphi \left( D_{8}\varphi^3 \Phi^2 + D_{9}\varphi \Phi^2 + D_{10}\varphi \right)  \right.\nonumber\\ 
&\left. + D_{11}\Phi^2 \varphi^4 + D_{12}\Phi^2 \varphi^2 +D_{13} \varphi^2 +D_{14} \Phi^2  \right.\nonumber\\ 
&\left. +  \chi \left( D_{15}\Phi'' + D_{16}\Phi'' \varphi \Phi + D_{17}\Delta \Phi + D_{18}\Delta \Phi \varphi \Phi +D_{19} \varphi'' + D_{20}\varphi'' \varphi \Phi +D_{21} \Delta \varphi +D_{22} \Delta \varphi \varphi \Phi \right) + D_{23}\chi^2  \right].\nonumber
\end{align}

After solving for all coefficients $D_i$, applying the field transformations

\begin{align}\label{eq:field_trafo_Phi_phi_chi}
\Phi &\longrightarrow  A_1 \pi + \ghost + \Mp^{-1}  \xi,\\
\varphi &\longrightarrow \Mp \pi - \Mp \ghost + A_2 \xi,\\
\chi &\longrightarrow A_3 \pi - \ghost + \Mp^{-1} \xi,
\end{align}

and setting $D_{23} = -m^2 \Mp^2$, we find by checking the relative signs of all kinetic terms that $\ghost$ is the ghost mode with the mass
\begin{align}\label{eq:mass_ghost}
m^2_{\ghost} \Mp^2 = C_{44} - m_\varphi^2 \Mp^2 - m^2 \Mp^2,
\end{align}

and $\pi$ and $\xi$ describe the helicity-0 mode and the matter field, respectively, with masses
\begin{align}\label{eq:mass_helicity0-and-matter}
m^2_{\pi} \Mp^2 &= C_{43} \Mp^2 + \frac{\left(\Mp^2 \left(m^2-C_{43}\right)+C_{44}\right)^2}{4C_{44}}-\frac{\left(\Mp^2 \left(C_{43}+m^2\right)+C_{44}\right)^2}{4m^2 \Mp^2}\\
&= m^2_{\ghost} \Mp^2 \left( -1 + \frac{1}{4} m^2_{\ghost} \left( C_{44}^{-1}\Mp^2 - m^{-2}\right) \right),\\
m^2_{\xi} &= \Mp^{-2}\left( C_{44} - m^2 \Mp^2 \right) + \frac{1}{C_{43} \Mp^4} \left( C_{44} + m^2 \Mp^2 \right)^2\\
&=\frac{m^2_{\ghost} m^2 \Mp^2 - C_{44} \left( m^2_{\ghost} + 4m^2 \right)}{C_{44} - \left( m^2_{\ghost} + m^2\right) \Mp^2} .
\end{align}

We observe that all masses are mainly determined by the coefficient $C_{44}$. In our favored parameter region that satisfies Eq. (\ref{eq:stability_curve}) and $\alpha_1 = \mathcal O (1)$ we find $C_{44} \gg m^2 \Mp^2$. Hence, if $m_\varphi^2$ is small (which we will also assume later for the analysis of the vacuum decay) then Eq. (\ref{eq:mass_ghost}) indicates a positive $m^2_{\ghost}$ but tachyonic scalars $\pi$ and $\xi$. This is not surprising as we have already seen the existence of a tachyon in the vacuum case. With an additional coupling to a new, even non-tachyonic, scalar field the tachyonic instability can leak into all other mass terms. However, this does not render the theory more dangerous and rather tells us that the decay processes in our theory of massive gravity with an additional scalar field can be described by the equivalent setting of one ghost and two tachyonic fields.

\subsection{Strong coupling scale of the theory\label{sub:Cutoff-of-the-theory}}

The constraint that removes the BD ghost in a non-linear theory of
a massive graviton automatically removes all interactions that are
suppressed by scales $\Lambda<\Lambda_{3}$ with
\begin{equation}
\Lambda_{\lambda}\equiv\left(M_{\text{P}}m^{\lambda-1}\right)^{1/\lambda}.
\end{equation}
All other non-linear theories that reduce to FP at the linear level contain
terms suppressed by $\Lambda_{5}$. However, this does not necessarily
hold for theories that do not reduce to the FP theory at the linear level.

We can find the cutoff scale of HMG by using the expansion of the mass term (\ref{eq:pot_2nd})
and introducing the St\"{u}ckelberg fields
\begin{equation}
\delta g_{\mu\nu}\longrightarrow h_{\mu\nu}+\partial_{\mu}A_{\nu}+\partial_{\nu}A_{\mu},
\end{equation}
and, subsequently,
\begin{equation}
A_{\mu}\longrightarrow A_{\mu}+\partial_{\mu}\phi.
\end{equation}
This decomposition into the three helicity modes allows us to read
off the energy scales with which all single interactions are suppressed
(see, e.g., Refs. \cite{Hinterbichler2011,deRham2014}). For this we
need to canonically normalize all the modes through the rescaling
\begin{align}
h_{\mu\nu}\longrightarrow & \frac{2}{M_{\text{P}}}h_{\mu\nu},\\
A_{\mu}\longrightarrow & \frac{2}{mM_{\text{P}}}A_{\mu},\\
\phi\longrightarrow & \frac{2}{m^{2}M_{\text{P}}}\phi.
\end{align}
One now finds that the interactions in HMG that are suppressed by
the smallest scale are of the type
\begin{equation}
\propto\frac{\bar{\alpha}_{i}}{M_{\text{P}}m^{4}}\left(\partial\partial\phi\right)^{3}=\frac{\bar{\alpha}_{i}}{\Lambda_{5}^{5}}\left(\partial\partial\phi\right)^{3},
\end{equation}
and correspond to the cutoff scale $\Lambda_{5}$.

\section{Quantum instability}

\subsection{Most dominant interaction terms}

The quantum stability depends on the scattering
between the ghost $\ghost$ and the matter field $\xi$. In order to compute the scattering amplitude
we move to Fourier space and introduce $k_{\ghostsmall}$ and $k_{\xi}$
for the momenta of $\ghost$ and $\xi$, respectively. The
final action of the interaction between these two fields is rather
complicated. In general, the interaction terms contain derivatives
of both fields that describe the so-called derivative interactions, which,
thus, have momentum-dependent vertices. It is exactly this type of interaction
that might be dangerous since the scattering amplitude requires an integration
over the entire phase space of the initial and final states of the
fields. Therefore, all derivative interactions lead to UV divergent
terms $\propto k^{\alpha}$ with $\alpha\in\mathbb{R}^{+}$. Even
though such derivative interactions exist in the Standard Model (SM),
this problem is usually solved by introducing counter terms which
regularize the divergent parts. In our case, we require a Lorentz
violation to cut the integral over the phase space.

If the integral of the phase space is cut at some energy level due to some new Lorentz breaking operators, then
the decay rate might not necessarily be dominated by the UV behavior
anymore. As seen in Sect. \ref{sub:Cutoff-of-the-theory}, the cutoff
of the EFT is much below the Planck scale. Depending on the mass of
the graviton, terms with a lower number of derivatives could then
become dominant. At the end of this section we will, however, find
that these types of interactions are indeed less important.

From now on, we will need to only focus on the interactions with the highest
number of derivatives where we are allowed to assume that both $k_{\ghostsmall}$
and $k_{\xi}$ are of the same order since the cutoff scales above which
Lorentz invariance is broken are equivalent for both momenta. Even though one can directly see from the action (\ref{eq:action_Phi_phi_chi}) that there are many different types of derivative interations, most of them are suppressed by powers of $\Mp^{-1}$ or $m_\varphi$. We find the Lagrangian corresponding to the most dangerous process to be
\begin{equation}\label{eq:dominant_interaction1}
\mathcal{L}^{\text{dom}}\simeq \frac{\left(c_2^2+4 c_2 (4 c_3+3 c_4+1)-8 \left(2 c_3^2+2 c_3 c_4+c_4 (2 c_4-1)\right)\right)^4 m^6}{32 m_\varphi^6 \Mp^2 (c_2+2 c_4)^5} \; \ghost^2\; \xi^3 \;\partial_\mu \partial^\mu \xi.
\end{equation}
For the analysis of the vacuum decay it will be useful to apply the transformation $\ghost \longrightarrow \Mp^{-1} \ghost$ to obtain the same dimensions for both $\ghost$ and $\xi$. Finally, the interaction in Fourier-space becomes
\begin{equation}\label{eq:dominant_interaction}
\mathcal{L}^{\text{dom}}\simeq \frac{\left(c_2^2+4 c_2 (4 c_3+3 c_4+1)-8 \left(2 c_3^2+2 c_3 c_4+c_4 (2 c_4-1)\right)\right)^4 m^6}{32 m_\varphi^6 \Mp^4 (c_2+2 c_4)^5} k_\xi^2 \; \ghost^2\; \xi^4.
\end{equation}

Note that this interaction arises from a matter sector, which, as in
many other theories of modified gravity, couples minimally to gravity.
Thus, this type of derivative interaction is not only a property of HMG
but rather occurs in a much broader class of theories, even beyond
massive gravity. Since we are studying the Lagrangian on-shell, the
exact term describing the most dominant interaction is, of course,
still model-dependent. Especially the occurrence of derivatives in
the potential term of the theory might lead to different results. However, we expect
that the qualitative results for HMG will still be valid for a huge
class of theories of modified gravity that introduce a ghost and
have a matter sector minimally coupled to gravity.

\subsection{Ghost decay}

\begin{figure}
\includegraphics[trim={0 0 0 0},clip,width=0.35\textwidth]{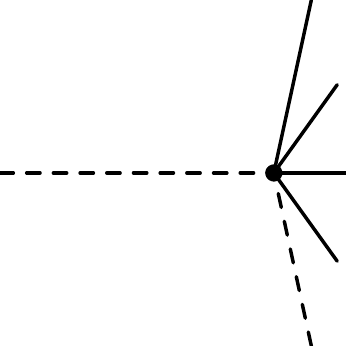}$\qquad$
\includegraphics[trim={7cm 0 0 0},clip,width=0.35\textwidth]{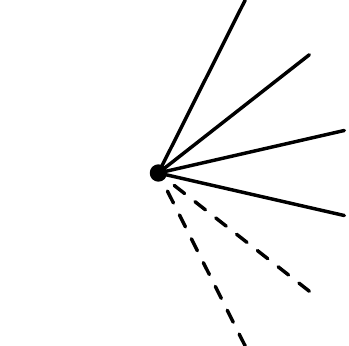}
\protect\protect\caption{\label{fig:FeynmanDiagrams} Left: Feynman diagram for the most dominant decay of a Boulware-Deser ghost particle (left dashed leg) into another ghost and four minimally coupled matter particles (solid legs). Right: Feynman diagram for the most dominant vacuum decay into ghost and matter paricles.}
\end{figure}

The total decay rate of the ghost particle is the sum of the decay rates from all possible decay channels.%
\footnote{Since the ghost is a boson, due to spin-statistics its production rate will be enhanced by a factor $1+n_{\vec p}$ depending on the occupation number of the final state. However, we are interested in the case in which the phase-space density of ghosts is negligible.}
As already discussed, the dominant contribution to the total decay rate comes from the process shown in Fig. \ref{fig:FeynmanDiagrams} (left) and, therefore, the rate can be very well approximated by
\begin{equation}\label{eq:ghost_decay_rate}
\Gamma_{\ghostsmall}=\frac{1}{2m_{\ghostsmall}}\int\prod_{f}\frac{\td ^{3}p_{f}}{\left(2\pi\right)^{3}2E_{f}}\left|\mathcal{M}\right|^{2}\left(2\pi\right)^{4}\delta^{\left(4\right)}\left(p_{\ghostsmall}-\sum_{f}p_{f}\right).
\end{equation}
Here, $\mathcal{M}$ is the scattering amplitude, $m_{\ghostsmall}$ and $p_{\ghostsmall}$ are the mass and four-momentum of the ghost particle,
respectively,\footnote{Even though a (non-tachyonic) ghost is usually recognized as a field with negative
kinetic energy, its mass term does also carry an additional minus
sign \cite{Sbisa2014}.%
} and $E_{f}$ is the energy of a particle appearing in a
final decay product. The dominance of high momenta in the decay rate further justifies the high energy limit leading to Eq. (\ref{eq:dominant_interaction1}).

It is important to note that we have different dispersion relations for a ghost and
a standard field. While for
a ghost we have $E_{\ghostsmall}=-\sqrt{m_{\ghostsmall}^{2}+\vec{p}_{\ghostsmall}^{2}}$, the
dispersion relation for standard matter fields is $E_{\text{sm}}=\sqrt{m_{\text{sm}}^{2}+\vec{p}_{\text{sm}}^{2}}$.
Here $\vec{p}_{\ghostsmall}$ and $\vec{p}_{\text{sm}}$ are the spatial momenta
for the ghost and matter fields, respectively.

Since Eq. (\ref{eq:ghost_decay_rate}) contains an integral over the entire phase space of all decay products, the decay rate is usually expected to be infinitely large. As mentioned earlier, a LB allows us to cut the integral at $\Lambda_{\text{LB}}$, which is, in fact, the energy scale that determines the decay time.

We now need to find the $\mathcal{M}$ matrix that corresponds to
the derivative interaction between $\ghost$ and $\xi$ as
described by Eq. ($\ref{eq:dominant_interaction1}$). The derivatives yield two
powers from the vertex of the interaction. Furthermore, we multiply
the vertex by a factor of $3!$ as we can freely swap all lines
that correspond to $\xi$. Thus, the scattering amplitude from
the Feynman diagram shown in Fig. \ref{fig:FeynmanDiagrams} (left)
becomes
\begin{align}\label{eq:mmatrix2}
\mathcal{M}  =3!A\left(ip_{3}\right)\left(ip_{3}\right) =-3!A\eta^{\mu\nu}\left(p_{3}\right)_{\mu}\left(p_{3}\right)_{\nu},
\end{align}
where we have introduced
\begin{equation}
A\equiv\frac{\left(c_2^2+4 c_2 (4 c_3+3 c_4+1)-8 \left(2 c_3^2+2 c_3 c_4+c_4 (2 c_4-1)\right)\right)^4 m^6}{32 m_\varphi^6 \Mp^4 (c_2+2 c_4)^5}.
\end{equation}
In order to find an upper bound on the decay rate or, equivalently, a lower
bound on the decay time, we consider the worst-case scenario
in which the matter field is almost  massless.
Even though this will generally lead to higher decay rates, it is
still a good approximation as the decay will be dominant at energies
near the LI-violating cutoff scale. Assuming isotropy in the
decay process, i.e., $\td^{3}p_{f}=4\pi p_{f}^{2}\td p_{f}$, fixing the
angles between different vectors, and using the momentum conservation,
we finally obtain the differential decay rate,
\begin{equation}
\td \Gamma_{\ghostsmall}\simeq-\frac{18 A^{2}p_{2}p_{3}p_{4}p_{5}p_{6} m_{\xi}^4(2\pi)^4\delta^{\left(4\right)}\left(p_{\ghostsmall}-\sum_{f=1}^{5}p_{f}\right)}{\left(2\pi\right)^{10}m_{\ghostsmall}}.\label{eq:decayrate_fin}
\end{equation}
We are now able to perform the phase-space integral in Eq. (\ref{eq:ghost_decay_rate})
up to the cutoff scale $\Lambda_{\text{LB}}$, at which Lorentz breaking occurs, and obtain 
\begin{equation}
\Gamma_{\ghostsmall}\simeq\frac{3{A}^{2}m_{\xi}^4\Lambda_{\text{LB}}^{6}}{2\left(2\pi\right)^{10}m_{\ghostsmall}} + \mathcal O \left(  \Lambda_{\text{LB}}^5 \right).\label{eq:decay_rate_result}
\end{equation}
Note that $A$ contains the scale with which the tree-level interaction term (\ref{eq:dominant_interaction}) is suppressed. If one would consider contributions from loops then their vertices that are suppressed by $\Lambda_5$ might lower the scale with which the decay rate is suppressed down to $\Lambda_5$.\footnote{We thank Claudia de Rham for discussions on this aspect.}

As mentioned before, the decay rate (\ref{eq:decay_rate_result})
corresponds only to the scattering process that dominates in the
UV. The validity of this assumption is not obvious for low cutoff scales.
From Eqs. (\ref{eq:sol_Psi}) and (\ref{eq:sol_B}) we find that interactions
with less derivatives of $\varphi$ introduce additional factors
of $m^{2}M_{\text{P}}^{2}$ in $A$. From a power counting we find that
the corresponding decay rate $\tilde{\Gamma}_{\ghostsmall}$ behaves like
\begin{equation}
\tilde{\Gamma}_{\ghostsmall} \propto m^{4}M_{\text{P}}^{4}\Lambda_{\text{LB}}^{-8}\Gamma_{\ghostsmall}.
\end{equation}
Therefore, for all cutoff scales that satisfy $\Lambda_{\text{LB}}\apprge\sqrt{mM_{\text{P}}}\simeq10^{-2}\,\text{eV}$
(for $m=\mathcal{O}\left(H_{0}\right)$) we do not expect higher decay
rates. As we will see in the next section, this condition is always satisfied
for cutoff scales $\Lambdamax$ with which the
decay would happen on a timescale of the Hubble time.

\subsection{Vacuum decay}

Besides the decay of a ghost, the vacuum itself can also decay into two ghosts and additional matter particles. The Feynman diagram for the most dominant vacuum decay is shown in Fig. \ref{fig:FeynmanDiagrams} (right). The main contribution to the decay rate of the vacuum comes from the same vertex that we found for the ghost decay and, thus, we find

\begin{equation}\label{eq:vac_decay_rate}
\Gamma_{\vac}=\int\prod_{f_{\ghostsmall}}\frac{\td ^{3}p_{f_{\ghostsmall}}}{\left(2\pi\right)^{3}2E_{f_{\ghostsmall}}} \prod_{f}\frac{\td ^{3}p_{f}}{\left(2\pi\right)^{3}2E_{f}}\left|\mathcal{M}\right|^{2}\left(2\pi\right)^{4}\delta^{\left(4\right)}\left(\sum_{f_{\ghostsmall}} p_{f_{\ghostsmall}} + \sum_{f}p_{f}\right).
\end{equation}

After choosing the rest-frame of the ghost particle $f_1$, the $\mathcal{M}$ matrix is, up to a symmetry factor $2!$, similar to Eq. (\ref{eq:mmatrix2}) and, thus, Eq. (\ref{eq:vac_decay_rate}) reduces to
\begin{equation}
\Gamma_{\vac} = 2\Gamma_{\ghostsmall}.
\end{equation}

The total decay rate of the interaction described by Eq. (\ref{eq:dominant_interaction}) is, however, not simply the sum of all  decay rates $\Gamma_i$. The vacuum $|0\rangle$ is defined as the state without any excitations, which is not a stable state if the theory contains ghost fields. The particle production rate from the vacuum decay is, therefore, only a good approximation for an initial vacuum state and might become less trustable as the vacuum decays. For this reason and since the decay of the vacuum is not more dangerous than the decay of the ghost, we now focus on the ghost decay only.

\subsection{Numerical calculations}

The decay rate possesses some model dependencies.
A priori the graviton mass scale $m$ is a free parameter. However,
if HMG is to be regarded as a theory of modified gravity that is supposed to solve the
dark energy problem by providing self-accelerating solutions, then
$m$ should not be chosen arbitrarily. The mass parameter $m$ determines
the scale at which modifications to GR become important and is therefore expected
to be $\sim H_{0}$.

There is however an additional model dependency. By using the fit (\ref{eq:stability_curve}), we get 
\begin{equation}
A\propto 1/\left(c_{2}+2c_{4}\right)^5\simeq\left(\frac{45}{26-120\bar{\alpha}_{1}}\right)^5,
\end{equation}
which, by tuning
$\bar{\alpha}_{1}$, might diverge, leading to an infinite decay rate. As discussed previously, this limit corresponds to a strong coupling of the matter and ghost mode, and thus, the perturbative approach breaks down.

\begin{figure}
\includegraphics[width=0.45\textwidth]{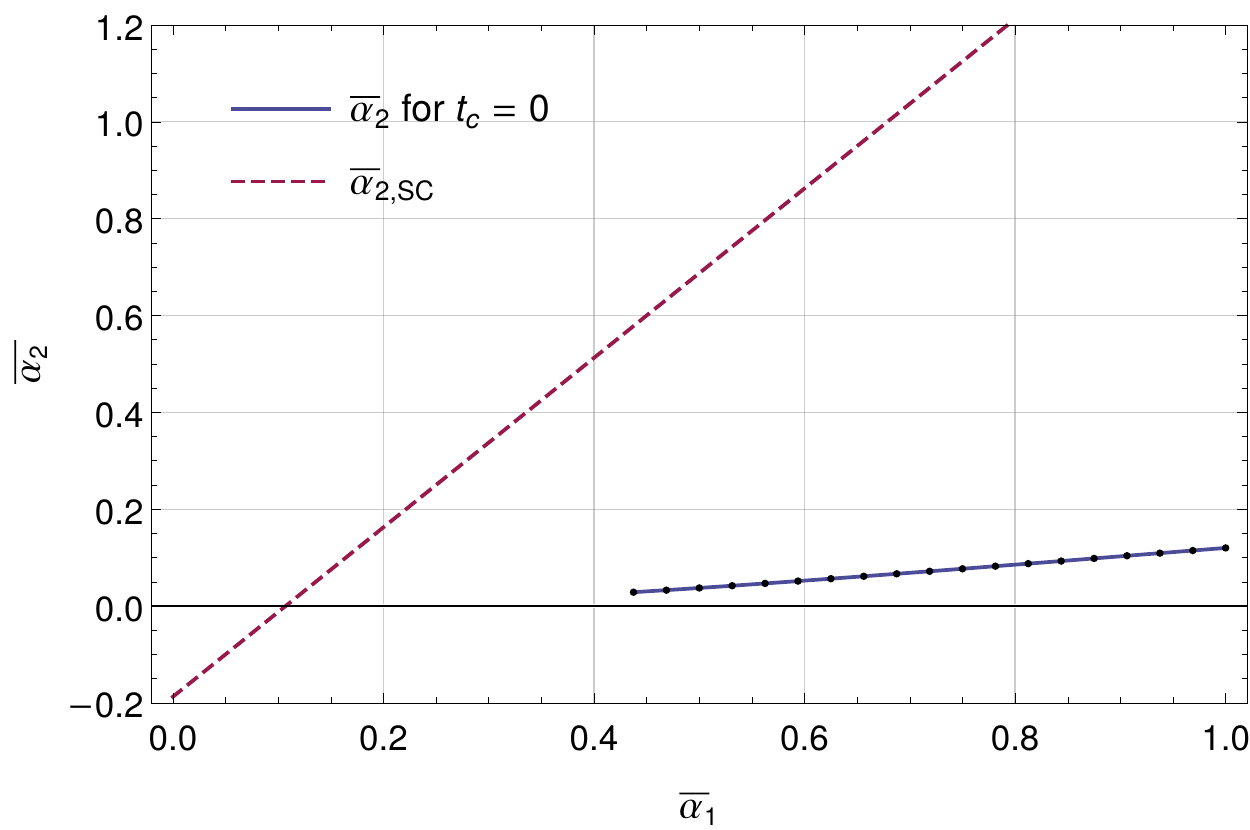}$\qquad$
\includegraphics[width=0.45\textwidth]{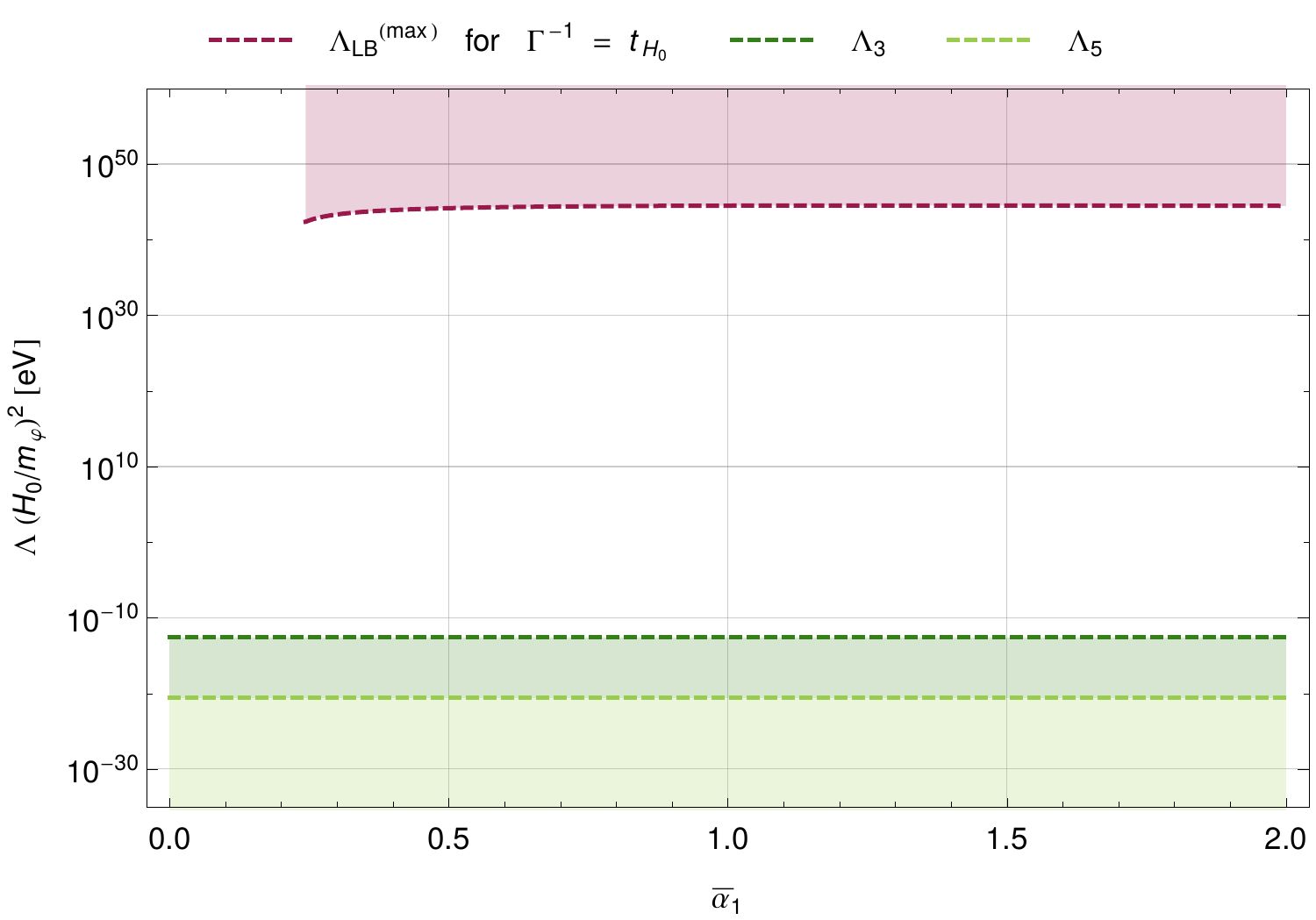}
\protect\protect\caption{\label{fig:stability_cutoff} Left: Constraints on the parameter space of HMG. The blue solid line indicates
the region in which Eq. (\ref{eq:stability_curve}) is satisfied, i.e., the timescale of the instability at the classical level to develop is maximized. Models on
the red dashed line satisfy $c_{2}+2c_{4}=0$, indicating the strong coupling regime. Right: Numerical results for the upper bound on the Lorentz-breaking cutoff scale, $\Lambdamax$
(red dashed line) corresponding to a decay time of the order of
the Hubble time $H_{0}^{-1}$. As indicated by the lowest
and second-lowest dashed lines, denoting $\Lambda_{5}$ and $\Lambda_{3}$,
respectively, the LB cutoff scale can be much larger than the strong coupling scale of the
EFT.}
\end{figure}

Since the classical background should also be unstable if the
vacuum decays at tree level, we do not expect to find stable classical
backgrounds for
$\bar{\alpha}_{1}\simeq 13/60$.
For a cross-check, we determine the parameter region that maximizes the timescale of the classical instability to develop. As shown in Fig. \ref{fig:stability_cutoff} (left panel),
we find that the results of the background analysis indeed agree
with this constraint.

The viability of the theory depends on the decay time of the most
dominant scattering process, and requires $\Gamma^{-1}\apprge H_{0}^{-1}$,
which sets an upper bound on the scale $\Lambda_{\text{LB}}$. For a graviton mass $m=\mathcal{O}\left(H_{0}\right)$ and $\bar{\alpha}_{1}=\mathcal{O}\left(1\right)$, where we approximately find $m_\xi\simeq m_{\ghostsmall}\simeq m$, we can estimate
the order of magnitude for the upper bound $\Lambdamax$,

\begin{equation}
\Lambda_{\text{LB}}\lesssim\Lambdamax \equiv\left( \frac{ 2 \left(2\pi\right)^{10} m m_{\ghostsmall}}{ 3 A^2 m_\xi^4}\right)^{1/6} = \mathcal O \left( \left( \frac{m_\varphi^{12} \Mp^8}{m^{14}} \right)^{1/6} \right).\label{eq:lambda_upper_bound}
\end{equation}
This gives us an upper limit that is much above the cutoff scale of the theory.
For more accurate numerical results see Fig. \ref{fig:stability_cutoff} (right panel). In the limit $m_\varphi \rightarrow 0$, the amplitude $A$ diverges and indicates an infinitely large decay rate. However, in this limit the Lagrangian of the interaction (\ref{eq:dominant_interaction}) would enter a strongly coupled regime and our perturbative ansatz would not be trustable anymore. Nevertheless, even considering extremely small masses $m_\varphi \simeq m=\mathcal O (H_0)$ would lead to $\Lambdamax > \Mp$.

Even though this LB cutoff scale $\Lambdamax$ is much above the strong coupling scale
that was found in Sect. \ref{sub:Cutoff-of-the-theory},
all of our results are still trustable and should be taken seriously. In
fact, the decay products can reach energies near $\Lambdamax$.
In addition, we should expect the decay processes
to occur even above $\Lambda_{\text{EFT}}$ (which is $\Lambda_5$ for our massive gravity theory). We should however note that it could indeed
be possible that energies above $\Lambda_{\text{EFT}}$ would lead to new
interactions that dominate and result in much larger decay rates, depending on the underlying new physics at energies above the EFT cutoff scale.

\subsection{Comparison to observations}

To date, no high-energy physics experiments have found any signals for the violation of Lorentz invariance, which may seem to indicate that Lorentz-violating operators, if exist, play a role only at very high energies, perhaps even above the Planck scale. Even though our results are compatible with this  conclusion, a LB at much smaller energy scales but above $\Lambda_3$ would nevertheless be allowed.


Even though the arguments above require some speculation about the
UV-completed physics, there is a more profound reason why it is
not surprising to find no LV at higher energies. As recently pointed
out in Ref. \cite{Afshordi2015}, most of the operators that break
LI lead to a strong coupling already above energies of $\mathcal{O}\left(\text{meV}\right)$. It has been conjectured that the strongly coupled degree of freedom (in our case the one that leads to a LV) effectively decouples
from the high-energy theory and can therefore not be observed
yet \cite{Afshordi2015}. Similar problems appear in QCD (confinement) and massive gravity
(Vainshtein screening).

Fortunately, there are possibilities to indirectly detect a breaking
of LI that stabilizes the vacuum decay. For the decay products it is most likely to have
energies of order $\Lambda_\text{LB}$, even if $\Lambda_\text{LB}\gg\Lambda_{\text{EFT}}$.
As long as they do not scatter at these energy levels, they can still
consistently be described by our EFT. A direct observation of these
decay products could then hint towards a breaking of LI. If one assumes a LB above $\sim 1 \text{MeV}$ then one could search for observable effects such as peaks in the gamma-ray background, along the lines of the studies in Ref~\cite{Cline2003}. However, the background flux is not well constrained yet for all (especially higher) energies.

\section{Summary and conclusions}

In this work, we have discussed the influence of a ghost on the viability
of an EFT by considering the violation of Lorentz invariance above certain energy scales in a particular theory of modified gravity describing
a massive graviton with an additional Boulware-Deser ghost, which we called haunted massive gravity (HMG). Even though
we do not believe that our HMG model is able to play a
major role in the class of theories of modified gravity attempting
to explain, e.g., the late-time acceleration of our Universe, we do expect that
its quantum properties can be mapped onto a huge class of other theories
of gravity that also introduce an Ostrogradski ghost.

In contrast to simple toy models with a canonical scalar field interacting
with a ghost, we have found a decay rate that does not scale as $\Lambda_{\text{LB}}^{2}$, where $\Lambda_{\text{LB}}$ denotes the energy scale above which Lorentz invariance is broken, or $\Lambda_{\text{LB}}^{8}$ if one assumes the simplest interaction with a graviton; the decay rate scales, instead, as $\Lambda_{\text{LB}}^{6}$. The origin of this difference
lies in the different dominating scattering processes involved. If the
ghost mass is of the order of the Hubble parameter $H_{0}$, which is expected
for theories of modified gravity that provide solutions
to the dark energy problem, then the upper bound on the cutoff scale at
which LB has to occur is allowed to be extremely high and could even be above the Planck scale.

Finally, with HMG we have found an example of a massive gravity theory which
allows for dynamical, and even self-accelerating, FLRW solutions
with a flat reference metric, contrary to the ghost-free dRGT theory. Furthermore, we obtained a parameter region in which both free parameters of the theory are of $\mathcal O \left( 1 \right)$ and maximizes the timescale on which the classical instability is suppressed to obtain a viable cosmological solution. This is indeed surprising as one might expect that a ghost that is present at the background level (which is required in order to obtain dynamical FLRW solutions) will automatically destabilize the theory. We have however studied only the background solutions, and one should therefore note that it is very likely that the cosmological perturbations
would be classically unstable, although it is not obvious with which timescale this instability is suppressed. Furthermore, it might also be possible that
quantum loops would render the theory unviable due to interactions that could
theoretically be much more dangerous than the tree-level interactions which
we have studied in this work; we leave the investigation of these questions for future work.

In general, ghosts are potentially dangerous and can rule
out a theory if the quantum behavior is not under control. However,
if one accepts the possibility of Lorentz-violating physics above
the cutoff of the theory, then all these theories should be studied
carefully and might be acceptable and well behaved.

\begin{acknowledgments}
We are grateful to Arthur Hebecker for initial discussions. We would also like to thank Claudia de Rham, Florian F\"{u}hrer, Johannes Noller, Sabir Ramazanov, Javier Rubio, Angnis Schmidt-May, Adam R. Solomon, and Mikael von Strauss for useful comments and suggestions. We are also grateful to the anonymous referee for important and helpful comments on a previous version of the manuscript. Y.A., L.A., F.K., and H.N. acknowledge support from DFG through the TRR33 project ``The
Dark Universe.'' F.K. is also supported by the Landesgraduiertenf\"{o}rderung
(LGFG) through the Graduate College ``Astrophysics of Fundamental
Probes of Gravity.'' H.N. also acknowledges financial support from DAAD through the program ``Forschungsstipendium f\"{u}r Doktoranden und Nachwuchswissenschaftler.''
Significant parts of the tensor algebra benefited from the package
xAct for Mathematica \cite{xAct}.
\end{acknowledgments}

\newpage

\appendix

\section{Detailed expressions for decoupling of the helicity-0 mode, the ghost and the matter field}\label{app_dec_three_dof}

The coefficients $C_i$ corresponding to the action (\ref{eq:action_Phi_phi}) read

\begin{align}\label{coeff_Ci_Phi_phi}
&C_{1} &=\,& 0,& \quad &C_{2} &=\,& 0,&\quad &C_{3} &=\,& -\frac{16\Mp^2}{\left(c_2+2c_4\right) m^2} ,&\\
 &C_{4} &=\,& -\frac{4}{\left(c_2+2c_4\right)m^2 \Mp^2},&\quad &C_{5} &=\,& 0,&\quad &C_{6} &=\,& 0 ,&\nonumber\\
 &C_{7} &=\,& 0,&\quad  &C_{8} &=\,& 0,&\quad &C_{9} &=\,& -\frac{16 \Mp^2}{\left(c_2+2c_4\right) m^2},&\nonumber\\
&C_{10} &=\,& -\frac{4}{\left(c_2+2c_4\right) m^2 \Mp^2} ,&\quad  &C_{11} &=\,& 0,&\quad  &C_{12} &=\,& 0,&\nonumber\\
&C_{13} &=\,& -\frac{16}{\left(c_2+2c_4\right)m^2},&\quad &C_{14} &=\,& \frac{16}{\left(c_2+2c_4\right) m^2} ,&\quad  &C_{15} &=\,& -\frac{16}{\left(c_2+2c_4\right)m^2},&\nonumber\\
 &C_{16} &=\,& 0,&\quad &C_{17} &=\,& 0 ,&\quad  &C_{18} &=\,& \frac{32\Mp^2}{\left(c_2+2c_4\right)m^2} ,&\nonumber\\
 &C_{19} &=\,&\frac{8}{\left(c_2+2c_4\right)m^2\Mp^2} ,&\quad  &C_{20} &=\,& \frac{16}{\left(c_2+2c_4\right)m^2},&\quad &C_{21} &=\,& 0 ,&\nonumber\\
 &C_{22} &=\,& 0 ,&\quad  &C_{23} &=\,& 0,&\quad  &C_{24} &=\,& 0,&\nonumber\\
&C_{25} &=\,& 0 ,&\quad  &C_{26} &=\,& 0 ,&\quad  &C_{27} &=\,& 0,&\nonumber\\
 &C_{28} &=\,& 0,&\quad &C_{29} &=\,& 0 ,&\quad  &C_{30} &=\,& 0 ,&\nonumber\\
 &C_{31} &=\,& -\frac{16m_\varphi^2}{\left(c_2+2c_4\right)m^2},&\quad  &C_{32} &=\,& \frac{8\Mp^2 \left(c_2-2c_3-c_4\right)}{c_2+2c_4},&\quad &C_{33} &=\,& -\frac{8m_\varphi^2}{\left(c_2+2c_4\right)^2 \Mp^2} ,&\nonumber\\
 &C_{34} &=\,& \frac{2\left(c_2 -4c_3 - 4c_4\right)}{c_2 + 2c_4} ,&\quad  &C_{35} &=\,& -1 ,&\quad  &C_{36} &=\,& \frac{16 m_\varphi^2}{\left(c_2+2c_4\right)m^2} ,&\nonumber\\
&C_{37} &=\,& -\frac{8\left(c_2 - 2c_3 - c_4\right)}{c_2 + 2c_4}  ,&\quad  &C_{38} &=\,& \frac{8m_\varphi^2}{\left(c_2+2c_4\right) m^2 \Mp^2} ,&\quad  &C_{39} &=\,& -\frac{2\left(c_2 - 4 c_3 - 4c_4\right)}{c_2+2c_4},&\nonumber\\
 &C_{40} &=\,& 1,&\quad &C_{41} &=\,& -\frac{4m_\varphi^4}{\left(c_2+2c_4\right)m^2\Mp^2} ,&\quad  &C_{42} &=\,& \frac{2\left(c_2-4c_3-4c_4\right)m_\varphi^2}{c_2 + 2c_4} ,&\nonumber\\
 &C_{43} &=\,& -m_\varphi^2,&\quad  &C_{44} &=\,& \frac{\left(c_2^2 + 4c_2\left(4c_3 + 3c_4\right)-16\left(c_3^2+c_3c_4+c_4^2\right)\right)m^2 \Mp^2}{4\left(c_2+2c_4\right)}.\span\span\span\span&\quad&&&\nonumber
\end{align}
After integrating out the auxiliary field $\chi$ in Eq. (\ref{eq:action_Phi_phi_chi}), the comparison of the resulting action with the original one (\ref{eq:action_Phi_phi}) provides a set of equations that can be solved with

\begin{align}
D_{i} &= C_i\;\text{ for }\; 1\leq i \leq 14,\\
D_{15} &= \mp 2 \sqrt{-C_3 D_{23}},\\
D_{16} &= D_{18} = D_{19} = D_{21} = 0,\\
D_{17} &= \mp \frac{2C_{13}\sqrt{-C_{3}D_{23}}}{C_{14}},\\
D_{20} &= \mp C_{15}\sqrt{\frac{D_{23}}{C_3}},\\
D_{22} &= \mp C_{14}\sqrt{\frac{D_{23}}{C_3}},
\end{align}
if the following contraints are fulfilled:
\begin{align}
C_{15}^2 &= 4C_3 C_4,\\
C_{13}^2 C_3 &= D_{14}^2 C_9,\\
4C_3 C_{10} &= C_{14}^2,\\
C_{14}C_{18} &= 2C_3 C_{13},\\
2C_3 C_{19} &= C_{14}C_{15},\\
C_{13}C_{15} &= C_{14}C_{20}.
\end{align}

All of them are indeed satisfied for HMG.
\par
For the transformations given in Eq. (\ref{eq:field_trafo_Phi_phi_chi}) and the choice $D_{23} = -m^2 \Mp^2$ we find that the mass matrix is diagonalized if
\begin{align}
2 A_1 C_{44} +2 A_3 m^2 \Mp^2-2 C_{43} \Mp^2&=0,\\
-2 A_2 C_{43}\Mp+\frac{2 C_{44}}{\Mp}+2 m^2\Mp&=0,\\
\frac{2 A_1 C_{44} }{\Mp}+2 A_2 C_{43} \Mp-2 A_3 m^2  \Mp &=0.
\end{align}

\bibliographystyle{apsrev}
\bibliography{references_FK,references_YA}

\end{document}